\documentclass[a4paper,11pt]{article}

\pdfoutput=1
\usepackage{jcappub}

\usepackage{xcolor} 
\usepackage{hyperref}
\usepackage{listings}
\usepackage{mathtools}
\usepackage{amsmath}
\usepackage{amsfonts}
\usepackage{amssymb}
\usepackage{bbold}
\usepackage{epsfig}
\usepackage{graphicx}
\usepackage{bm}
\usepackage{array}
\usepackage{hyperref}
\usepackage{listings}
\usepackage{color}
\usepackage{float}
\usepackage[normalem]{ulem}
\usepackage{placeins}
\usepackage{bbold}

\newcommand{\exclude}[1]{}

\usepackage{tikz,xcolor,hyperref}

\definecolor{lime}{HTML}{A6CE39}
\DeclareRobustCommand{\orcidicon}{\hspace{-1mm}
 \begin{tikzpicture}
 \draw[lime, fill=lime] (0,0) 
 circle [radius=0.16] 
 node[white] {{\fontfamily{qag}\selectfont \tiny \,ID}};
 \draw[white, fill=white] (-0.0525,0.095) 
 circle [radius=0.007];
 \end{tikzpicture}
 \hspace{-3mm}
}

\foreach \x in {A, ..., Z}{\expandafter\xdef\csname orcid\x\endcsname{\noexpand\href{https://orcid.org/\csname orcidauthor\x\endcsname}
 {\noexpand\orcidicon}}
}

\title{Neutrino quantum kinetics in a core-collapse supernova}

\author[a]{Shashank Shalgar\orcidB{}}

\author[a]{and Irene Tamborra\orcidC{}}

\affiliation[a]{Niels Bohr International Academy \& DARK, Niels Bohr Institute,\\University of Copenhagen, Blegdamsvej 17, 2100 Copenhagen, Denmark}

\emailAdd{shashank.shalgar@nbi.ku.dk}
\emailAdd{tamborra@nbi.ku.dk}

\abstract{Our understanding of neutrino flavor conversion in the supernova core is still preliminary, despite its likely relevance to the neutrino-driven supernova mechanism. We present multi-angle and multi-energy numerical simulations of neutrino quantum kinetics within a spherically symmetric shell 
in the proximity of the region of neutrino decoupling. 
We rely on inputs from a one-dimensional core-collapse supernova model with a mass of $18.6\ M_\odot$ and find that, at early post-bounce times ($t_{\mathrm pb} \lesssim 0.5$~s), no crossing is present in the angular distribution of the electron neutrino lepton number and flavor conversion is triggered by slow collective instabilities.
Angular crossings appear for $t_{\textrm{pb}} \gtrsim 0.5$~s and fast flavor conversion leads to flavor equipartition, with the spectral energy distribution of $\nu_{e}$ ($\bar{\nu}_{e}$) and $\nu_{x}$ ($\bar{\nu}_{x}$) becoming comparable. Notably, flavor equipartition is not a generic outcome of fast flavor conversion,  rather it is a consequence of the relatively similar properties of neutrinos of different flavors characterizing the late accretion phase. Artificially tweaking the collision term to introduce an electron lepton number angular crossing for $t_{\mathrm{pb}} \lesssim 0.05$~s, we observe that flavor equipartition is not achieved. 
While our findings are restricted to a specific supernova model, and they only take into account the feedback of the neutrino background on the flavor conversion, they suggest a rich phenomenology in the supernova core as a function of the post-bounce time which needs to be further explored to assess its impact on the explosion mechanism. 
}

\begin{document}

\maketitle

\section{Introduction}
\label{sec:intro}
In neutrino dense astrophysical sources, flavor evolution is significantly affected by the coherent forward scattering (refraction) of neutrinos with each other~\cite{Pantaleone:1994ns,Langacker:1992xk}.
Neutrino self-interaction, however, makes the neutrino equations of motion nonlinear and the consequent flavor evolution displays a rich phenomenology that we are still far from understanding~\cite{Mirizzi:2015eza, Duan:2010bg, Chakraborty:2016yeg, Tamborra:2020cul,Richers:2022zug}.

The modification of neutrino flavor evolution due to refraction with other neutrinos can be broadly classified into slow~\cite{Duan:2005cp,Duan:2006an,Duan:2010bg}, fast~\cite{Sawyer:2005jk, Sawyer:2008zs, Sawyer:2015dsa, Chakraborty:2016lct, Izaguirre:2016gsx, Tamborra:2020cul,Richers:2022zug}, and collisional~\cite{Johns:2021qby}. Slow flavor evolution is differentiated from the fast one by the role of the vacuum frequency. In fact, fast flavor evolution can manifest in the limit of vanishing vacuum frequency and it is triggered by a crossing in the electron lepton number (ELN) angular distribution of neutrinos~\cite{Izaguirre:2016gsx,Chakraborty:2016lct,Morinaga:2021vmc,Fiorillo:2024bzm}, whereas slow flavor evolution requires a nonzero vacuum term~\cite{Duan:2006an,Hannestad:2006nj,Fiorillo:2023mze}. 
Fast flavor evolution, as the name suggests, occurs over very small time scales---the characteristic frequency being directly proportional to the number density of neutrinos.
 On the other hand, collisional instabilities should be triggered by neutrino collisions with the medium in the high-density region in the proximity of neutrino decoupling~\cite{Johns:2022yqy,Padilla-Gay:2022wck,Fiorillo:2023ajs,Lin:2022dek,Xiong:2022vsy}, but their relevance on flavor conversion physics within realistic astrophysical environments remains to be assessed~\cite{Nagakura:2023xhc,Shalgar:2023aca}.

Neutrinos play a crucial role in the core-collapse supernova mechanism~\cite{Colgate:1966ax,Bethe:1985sox}, as recently confirmed by multi-dimensional hydrodynamic simulations~\cite{Janka:2016fox,Muller:2020ard,Mezzacappa:2020oyq,Burrows:2020qrp}.
Whether neutrino flavor conversion, triggered by any of the flavor instabilities highlighted above, affects the core collapse mechanism is the subject of ongoing research~\cite{Ehring:2023lcd,Ehring:2023abs,Nagakura:2023mhr}. Preliminary work suggests that, if flavor equilibration should occur as a result of such instabilities, this would facilitate or hinder the supernova explosion according to the mass of the collapsing star~\cite{Ehring:2023abs}. Because of the technical challenges linked to the solution of the neutrino quantum kinetic equations (QKEs) and the vast difference in the characteristic scales typical of the source hydrodynamics and the ones entering the neutrino equations of motion, a self-consistent solution of this problem is not yet available~\cite{Tamborra:2020cul,Richers:2022zug,Johns:2023jjt,Johns:2024dbe}. 
Nevertheless, it is possible to search for general trends by solving the QKEs within a simulation shell, taking into account the interplay among flavor evolution, collisions, and advection. In this case, a quasi-steady state configuration is found in the aftermath of flavor conversion~\cite{Shalgar:2022rjj, Shalgar:2022lvv,Nagakura:2022xwe,Nagakura:2023mhr,Nagakura:2023xhc,Xiong:2024pue,Cornelius:2023eop,Shalgar:2019qwg}. 

An open question concerns whether flavor equipartition (i.e.~the angular and/or spectral distributions of the electron and non-electron flavors approach each other) is a general solution of the QKEs. While recent work seems to point in this direction~\cite{Xiong:2024tac,Xiong:2024pue,Martin:2021xyl,Zaizen:2023ihz,Zaizen:2022cik,Xiong:2023vcm,Grohs:2022fyq,Richers:2021xtf,Richers:2022bkd,Bhattacharyya:2020jpj,Bhattacharyya:2022eed,Wu:2021uvt}, the boundary conditions as well as the ELN distribution entering the solution of the QKEs could drastically affect the final flavor configuration~\cite{Cornelius:2023eop,Shalgar:2022rjj,Shalgar:2022lvv}. In this paper, we investigate the neutrino flavor evolution in spherical symmetry, relying on input from several post-bounce time snapshots within $1$~s of a one-dimensional hydrodynamic simulation of a core-collapse supernova with a mass of $18.6\ M_{\odot}$~\cite{SNarchive}. We find that, in the early accretion phase ($t_{\textrm{pb}} \lesssim 0.5$~s), flavor conversion is triggered by slow collective effects; while, in the late accretion phase ($t_{\textrm{pb}} \gtrsim 0.5$~s), ELN crossings in the angular distribution lead to flavor conversion, with resulting equipartition between the neutrino spectral energy distributions of $\nu_e$ and $\nu_x$ as well as $\bar\nu_e$ and $\bar\nu_x$. Yet, we confirm early findings~\cite{Shalgar:2022rjj, Shalgar:2022lvv,Cornelius:2023eop} showing that flavor equipartition is not a general flavor outcome for any given unstable flavor configuration.

This paper is organized as follows. Section~\ref{sec:QKE} introduces the neutrino QKEs and outlines the approach adopted to model the collision term as well as the simulation setup. In Sec.~\ref{sec:SNmodel}, we describe the main features of our benchmark one-dimensional supernova model, compute the neutrino angular distributions of all flavors employing a multi-energy transport scheme and illustrate as they evolve as functions of the post-bounce time. Our main findings are presented in Sec.~\ref{sec:results}.
Finally, we discuss our results in Sec.~\ref{sec:discussion} and conclude in Sec.~\ref{sec:conclusions}. Appendix~\ref{comparison} provides a comparison between  the neutrino number densities computed in this work and the ones obtained  from our benchmark hydrodynamical  simulation.

\section{Neutrino quantum kinetics in spherical symmetry}\label{sec:QKE}
In this section, we introduce the neutrino equations of motion. We also outline the numerical techniques adopted to solve the neutrino QKEs.

\subsection{Quantum kinetic equations}

\begin{figure}
\centering
\includegraphics[width=0.6\textwidth]{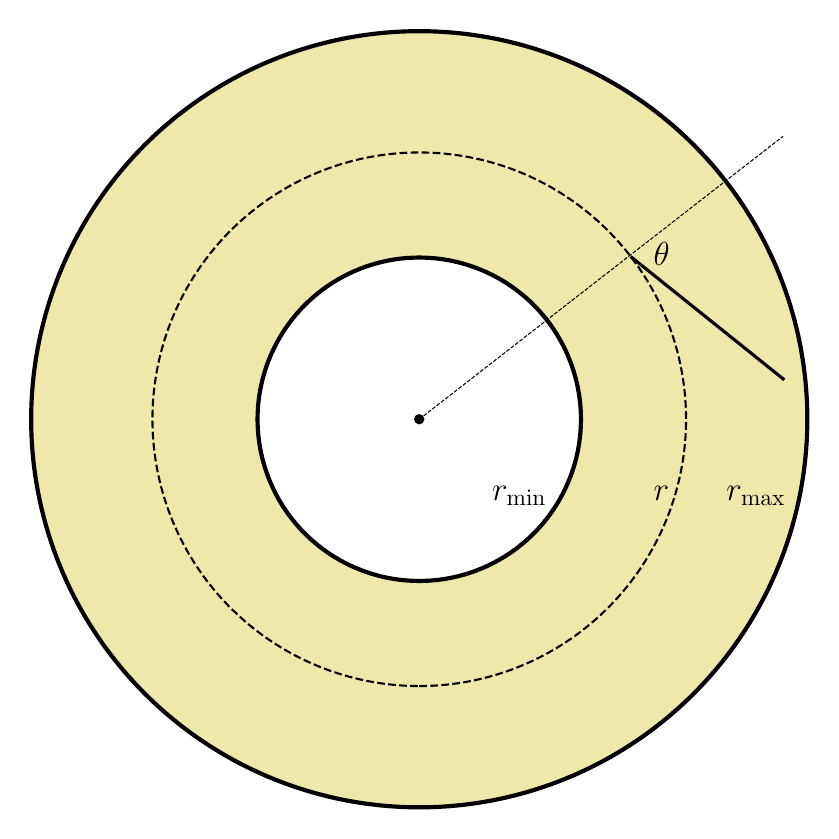}
\caption{Sketch of the simulation shell adopted to solve the QKEs. The polar angle $\theta$ is defined with respect to the radial direction and it is a function of radius ($r$) for any given trajectory. The radii $r_{\rm min}$ and $r_{\rm max}$ denote the innermost and outermost radii adopted in the simulation.}
\label{Fig0}
\end{figure}
For simplicity, we consider two neutrino flavors ($\nu_e$ and $\nu_x$ with $x = \mu, \tau$), and the corresponding antineutrinos ($\bar\nu_e$ and $\bar\nu_x$). We model our neutrino ensemble in terms of $2 \times 2$ density matrices for neutrinos and antineutrinos for each radial location ($r$), polar angle ($\theta$), and energy ($E$), as a function of time ($t$) and under the assumption of spherical symmetry, see Fig.~\ref{Fig0}~\footnote{Note that the polar angle $\theta$ is defined with respect to the radial direction at a given point and it is not the emission angle used in the early papers on neutrino-bulb model~\cite{Duan:2005cp,Duan:2006an,Duan:2010bg}.}.  
The diagonal elements of the density matrices ($\rho_{ii}$ with $i=e$ and $x$) represent the neutrino occupation number and are normalized such $\int_{-1}^{1}\rho(r,\cos\theta,E,t)dE~d\cos\theta$ gives the local neutrino number density of flavor $i$ at a given radius ($n_{\nu_i}$).

The equations of motion that describe the evolution of the (anti)neutrino density matrices are~\cite{Sigl:1992fn}:
\begin{eqnarray}
\label{eom1}
i\left(\frac{\partial}{\partial t} + \vec{v} \cdot \vec{\nabla}\right)\rho(r,\cos\theta,E,t) &=& [H(r,\cos\theta,E,t),\rho(r,\cos\theta,E,t)] + i\mathcal{C}[\rho] \ ,\\
\label{eom2}
i\left(\frac{\partial}{\partial t} + \vec{v} \cdot \vec{\nabla}\right)\bar{\rho}(r,\cos\theta,E,t) &=& [\bar{H}(r,\cos\theta,E,t),\bar{\rho}(r,\cos\theta,E,t)] + i\bar{\mathcal{C}}[\bar{\rho}]\ .
\end{eqnarray}
The total derivative on the left hand side of Eqs.~\ref{eom1} and \ref{eom2} consists of a partial derivative with respect to time and the advective term, which is in spherical symmetry 
\begin{eqnarray}
\vec{v} \cdot \vec{\nabla} = \cos\theta \frac{\partial}{\partial r} + \frac{\sin^{2}\theta}{r} \frac{\partial}{\partial \cos\theta}.
\end{eqnarray}
The advective term is such that, in the absence of any term other than the total derivative, neutrinos travel along a straight line. General relativistic effects are neglected in Eqs.~\ref{eom1} and \ref{eom2}, qualitatively they should not affect the interplay between neutrino flavor conversion and transport effects on small scales~\cite{Nagakura:2023mhr}.

On the right hand side of Eqs.~\ref{eom1} and \ref{eom2}, the commutator of the Hamiltonian and the density matrix takes into account neutrino flavor conversion physics. The Hamiltonian consists of three components: the vacuum, matter, and self-interaction terms. For the sake of simplicity, we choose to neglect the matter term and adopt an effective mixing angle:
\begin{equation}
H(r,\cos\theta,E,t) = H_{\textrm{vac}}(E) + H_{\nu\nu}(r,\cos\theta, t)\ \mathrm{and}\ 
\bar{H}(r,\cos\theta,E,t) = -H_{\textrm{vac}}(E) + H_{\nu\nu}(r,\cos\theta,t)\ .
\end{equation}
The vacuum and self-interaction terms are defined as
\begin{equation}
H_{\textrm{vac}}(E) = \frac{\omega_{\textrm{vac}}}{2}
\begin{pmatrix}
-\cos 2\vartheta_{\textrm{V}} & \sin 2\vartheta_{\textrm{V}} \\
\sin 2\vartheta_{\textrm{V}} & \cos 2 \vartheta_{\textrm{V}}
\end{pmatrix}\ ,\\
\end{equation}
\begin{eqnarray}
\label{eq:Hnunu}
		H_{\nu\nu}(r,\cos\theta,t) =  \sqrt{2} G_{\textrm{F}}  \zeta  \int_{-1}^{1}\int_{0}^{\infty} & & \left[\rho(r,\cos \theta^{\prime},E,t) - \bar{\rho}(r,\cos \theta^{\prime},E,t)\right] \nonumber \\
& \times &  (1-\cos\theta \cos\theta^{\prime}) dE ~d\cos\theta^{\prime}\ .
\end{eqnarray}
The vacuum frequency is $\omega_{\textrm{vac}} = {\Delta m^{2}}/{2 E}$ with $\Delta m^2 = 2.5 \times 10^{-3}$~eV$^2$,
and $\vartheta_{\textrm{V}} = 10^{-3}$~rad denotes the effective mixing angle. Although a small vacuum mixing angle is often adopted to mimic the effect of the matter background~\cite{Hannestad:2006nj},  whether flavor conversion is suppressed or not because of the large matter background~\cite{Esteban-Pretel:2008ovd,Dasgupta:2015iia,Abbar:2015fwa,Sigl:2021tmj}  should be subject of dedicated work for boundary problems like ours. In this work, we are only interested in exploring the flavor conversion outcome for different ELN configurations; however, we have tested that the quasi-steady state flavor configuration is unchanged for $\vartheta_{\textrm{V}} = 10^{-5}$~rad and $\vartheta_{\textrm{V}} = 10^{-6}$~rad (results not shown here). The self-interaction Hamiltonian, unlike the vacuum Hamiltonian, depends on $\theta$ and is obtained integrating over all momentum modes. The integration over the azimuthal angle results in a constant factor $2\pi$ that has been absorbed in the definition of the density matrices. On the other hand, the self-interaction Hamiltonian does not depend on energy. Following the attenuation method introduced in Ref.~\cite{Nagakura:2022kic}, we rescale the strength of neutrino self-interaction by a constant factor $\zeta=10^{-3}$. 
The choice of $\zeta$ is such that the self-interaction strength is always larger than $\omega$ at neutrino decoupling and the quasi-steady state flavor state configuration is likely not affected, but the time needed for reaching such a configuration changes with the computational time being reduced~\cite{Shalgar:2019qwg,Padilla-Gay:2020uxa}.

The collision term on the right hand side of Eqs.~\ref{eom1} and \ref{eom2}, represented by $\mathcal{C}$ and $\bar{\mathcal{C}}$, includes three contributions: the emission term ($\mathcal{C}^{\textrm{emit}}$), the absorption term ($\mathcal{C}^{\textrm{absorb}}$), and the direction changing term ($\mathcal{C}^{\textrm{dir-ch}}$)~\cite{1990Ap&SS.165...65R,Sigl:1992fn, Janka:2012wk, Mezzacappa:2020oyq}. For all flavors of neutrinos, pair production and Bremsstrahlung contribute to the emission and absorption terms, whereas the beta processes only contribute to electron-type neutrinos and antineutrinos.
The direction changing term is given by the neutral current interaction between neutrinos and nucleons, which are assumed to be elastic. The inelastic contribution is negligible for the electron type neutrinos and thus does not affect the presence of ELN crossings. The collision terms are defined as follows: 
\begin{eqnarray}
\label{coll1}
\mathcal{C}[\rho] &=& \mathcal{C}_{\textrm{emit}}(r,E) - \mathcal{C}_{\textrm{absorb}}(r,E) \odot \rho(r,\cos\theta,E) 
+ \cos\theta\ \mathcal{C}_{\textrm{ani}}(r,E) \int d\cos\theta^{\prime} \cos\theta^{\prime} \rho(r,\cos\theta^{\prime},E)
 \nonumber \\
 &+& \frac{\mathcal{C}_{\textrm{dir-ch}}(r,E)}{2}\int d \cos\theta^\prime \left[-\rho(r,\cos\theta,E)+\rho(r,\cos\theta^{\prime},E)\right] \ ;
\end{eqnarray}
we use the symbol $\odot$ to denote the elementwise multiplication:
\begin{eqnarray}
\mathcal{C}_{\textrm{absorb}}(E) \odot \rho(\cos\theta,E) &=& 
\begin{pmatrix}
 \mathcal{C}_{\textrm{absorb}}^{ee}(E) \rho_{ee}(\cos\theta,E) & \mathcal{C}_{\textrm{absorb}}^{ex}(E) \rho_{ex}(\cos\theta,E) \cr
 \mathcal{C}_{\textrm{absorb}}^{xe}(E) \rho_{xe}(\cos\theta,E) & \mathcal{C}_{\textrm{absorb}}^{xx}(E) \rho_{xx}(\cos\theta,E)
\end{pmatrix} \ . 
\end{eqnarray}
The emission and absorption terms are related by Kirchoff's law for each energy, 
${\mathcal{C}_{\textrm{emit}}(E)}/{\mathcal{C}_{\textrm{absorb}}(E)}= f_{\textrm{FD}}(E)$, 
 where $f^{\nu_{i}}_{\textrm{FD}}(E)={dn_{\nu_{i}}}/{dE}$ is the Fermi-Dirac energy distribution. An analogous expression holds for $\mathcal{C}[\bar\rho]$.
The terms $\mathcal{C}_{\textrm{ani}}$ and $\bar{\mathcal{C}}_{\textrm{ani}}$ are included out of completeness and take into account the anisotropic nature of direction changing neutral current interactions.
The collision terms used in this paper are the multi-energy extension of the ones presented in Appendix A of Ref.~\cite{Shalgar:2023aca}, which in turn are based on 
Refs.~\cite{thompsonthesis, OConnor:2014sgn}.

\subsection{Simulation setup}\label{sec:num}

In order to solve the QKEs, we use the Julia implementation of the Adam-Moultan multistep integrator, with adaptive order and adaptive step size for the time variable~\footnote{The length scale characteristic of our neutrino ensemble is not given by the inverse of the self-interaction potential, rather it is a combination of the collision term and the self-interaction potential.   We refer the interested reader to Sec.~VI and Appendix A of Ref.~\cite{Shalgar:2022rjj} and 
Sec.~IV and Appendix A of Ref.~\cite{Shalgar:2022lvv}}. Yet, it is important to ensure that the time step is much smaller than the inverse of the neutrino self-interaction strength; we do this by implementing an adaptive step size. (VCABM solver). 
The spatial derivatives in the advective term are calculated using the central difference method. Numerical convergence (test results not shown here) has been obtained by employing $75$ angle bins, $24$ energy bins for $E \in [0,50]$~MeV, and $150$ radial bins for $r \in [r_{\rm{min}}, r_{\rm{max}}]$ (with $r_{\rm{min}}$ ans $r_{\rm{max}}$ being the innermost and outermost radii of our simulation shell; see Fig.~\ref{Fig0}).

Our goal is to compute the ``quasi-steady state'' configuration achieved by the system~\footnote{A quasi-steady state configuration does not coincide with the solution of the boundary value problem obtained by setting the time derivative to zero. The quasi-steady state is also not approximately close to the solution of that boundary value problem. 
The reason for this counter-intuitive assertion is that, if Eqs.~\ref{eom1} and \ref{eom2} are evolved for a long enough time, $\rho_{ee}$, $\rho_{xx}$ and $|\rho_{ex}|$ reach a steady state, but the phase of $\rho_{ex}$ does not converge towards a steady state. This explains why we need to evolve the system as a function of time instead of attempting to find a solution to Eqs.~\ref{eom1} and \ref{eom2} setting the time derivative to zero.}. 
The latter should be independent of the initial configuration (cf.~Ref.~\cite{Cornelius:2023eop}), but the choice of the initial configuration does affect the efficiency of the numerical solution. Future work should be devoted to assess the putative existence of such unique quasi-steady state configuration, see e.g.~Ref.~\cite{Fiorillo:2024qbl}. As pointed out in Refs.~\cite{Shalgar:2022rjj,Shalgar:2022lvv}, the most efficient way to tackle this issue numerically is to first solve Eqs.~\ref{eom1} and \ref{eom2} for $H=\bar{H}=0$ to obtain the ``classical steady state'' solution (i.e., we solve the classical Boltzmann equation). We then use the classical steady state solution as the initial configuration to solve Eqs.~\ref{eom1} and \ref{eom2} for $H\neq 0$ and $\bar{H}\neq 0$. This approach has two advantages. First, it is easier to highlight the impact of flavor evolution; second, the initial classical steady state configuration is not too different from the final quasi-steady state. We stress, however, that the choice to use the classical solution as the initial configuration is not unique, in principle it might be possible to devise a better initial configuration.

For obtaining the classical steady-state solution as well as the quasi-steady state one, the radial range has to be chosen such that the neutrino spectral energy distributions coincide with Fermi-Dirac distributions at the minimum radius ($r_{\textrm{min}}$); whereas the maximum radius ($r_{\textrm{max}}$) is chosen such that the classical solution has negligible backward flux~\cite{Shalgar:2023aca}, see also Fig.~\ref{Fig0} and Table~\ref{table:rminrmax}. These conditions serve as boundary conditions for Eqs.~\ref{eom1} and \ref{eom2}. 
\begin{table}
\label{table:rminrmax}
\caption{Values of $r_{\textrm{min}}$ and $r_{\textrm{max}}$ chosen for each post-bounce time snapshot in order to capture the radial region of neutrino decoupling (see also Fig.~\ref{Fig0}).}
\centering
\begin{tabular}{|l|l|l|}
\hline
$t_{\rm{pb}}$ (s) & $r_{\textrm{min}}$ (km) & $r_{\textrm{max}}$ (km) \\
\hline
\hline
$0.05$ & 25 & 150 \\
\hline
$0.12$ & 25 & 100 \\
\hline
$0.25$ & 22 & 57 \\
\hline
$0.5$ & 20 & 35 \\
\hline
$0.75$ & 17 & 32 \\
\hline
$1$ & 16 & 31\\
\hline
\end{tabular}
\end{table}
The hydrodynamic and thermodynamic supernova properties employed to compute the classical and quasi-steady state configurations are extracted from a one-dimensional core-collapse supernova simulation whose features are illustrated in Sec.~\ref{sec:SNmodel}.

\section{Benchmark core-collapse supernova model}\label{sec:SNmodel}

We rely on the outputs of a one-dimensional hydrodynamic simulation of a $18.6\ M_\odot$ core-collapse supernova, with $1.4\ M_\odot$ gravitational mass and SFHo nuclear equation of state without muons~\cite{SNarchive}. In this model, the effects of proto-neutron star convection have been taken into account through a mixing-length approximation~\cite{Mirizzi:2015eza,2004cgps.book.....W}.

\begin{figure}
\centering
\includegraphics[width=0.98\textwidth]{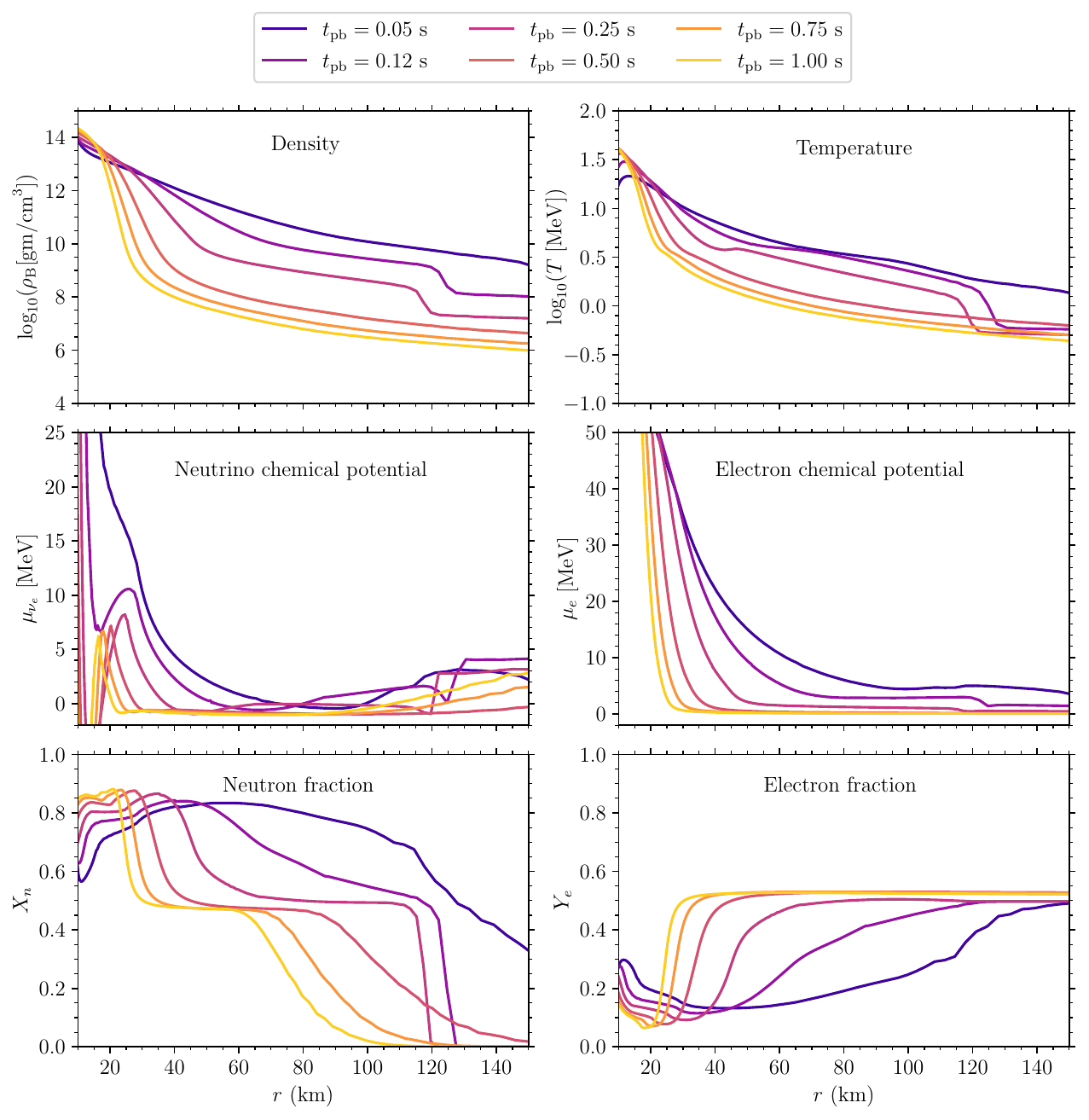}
\caption{Radial profiles of the characteristic quantities extracted from our core-collapse supernova model (see main text for details) at $t_{\rm pb} = 0.05$, $0.12$, $0.25$, $0.5$, $0.75$, and $1$~s. From top left to bottom right, each panel represents the radial evolution of the baryon number density, the electron temperature, the neutrino and electron chemical potentials, and the free neutron and electron fractions, respectively. }
\label{Fig0a}
\end{figure}

We focus on few selected post-bounce times ($t_{\rm pb} = 0.05$, $0.12$, $0.25$, $0.5$, $0.75$, and $1$~s) and rely on static hydrodynamic backgrounds and thermodynamical quantities for each time snapshot to compute the classical steady-state neutrino configuration. Figure~\ref{Fig0a} shows the radial evolution of the main supernova properties entering the collision term in Eqs~\ref{eom1} and \ref{eom2}. 

Since the neutrino angular distributions are not provided as an output of our benchmark hydrodynamic simulation, we compute them following the same procedure illustrated in Sec.~III of Ref.~\cite{Shalgar:2023aca}, i.e.~we set $r_{\rm min}$ and $r_{\max}$ as indicated in Table~\ref{table:rminrmax} for each snapshot and solve Eqs.~\ref{eom1} and \ref{eom2} for $H=\bar{H}=0$. 
 In contrast to Ref.~\cite{Shalgar:2023aca}, we compute the angular distributions of (anti)neutrinos solving Eqs.~\ref{eom1} and \ref{eom2} for multiple energy modes (see Sec.~\ref{sec:num}). Appendix~\ref{comparison} provides a comparison between the neutrino number densities computed through this approach and the ones obtained from the supernova hydrodynamic simulation.

\begin{figure}
\centering
\includegraphics[width=0.9\textwidth]{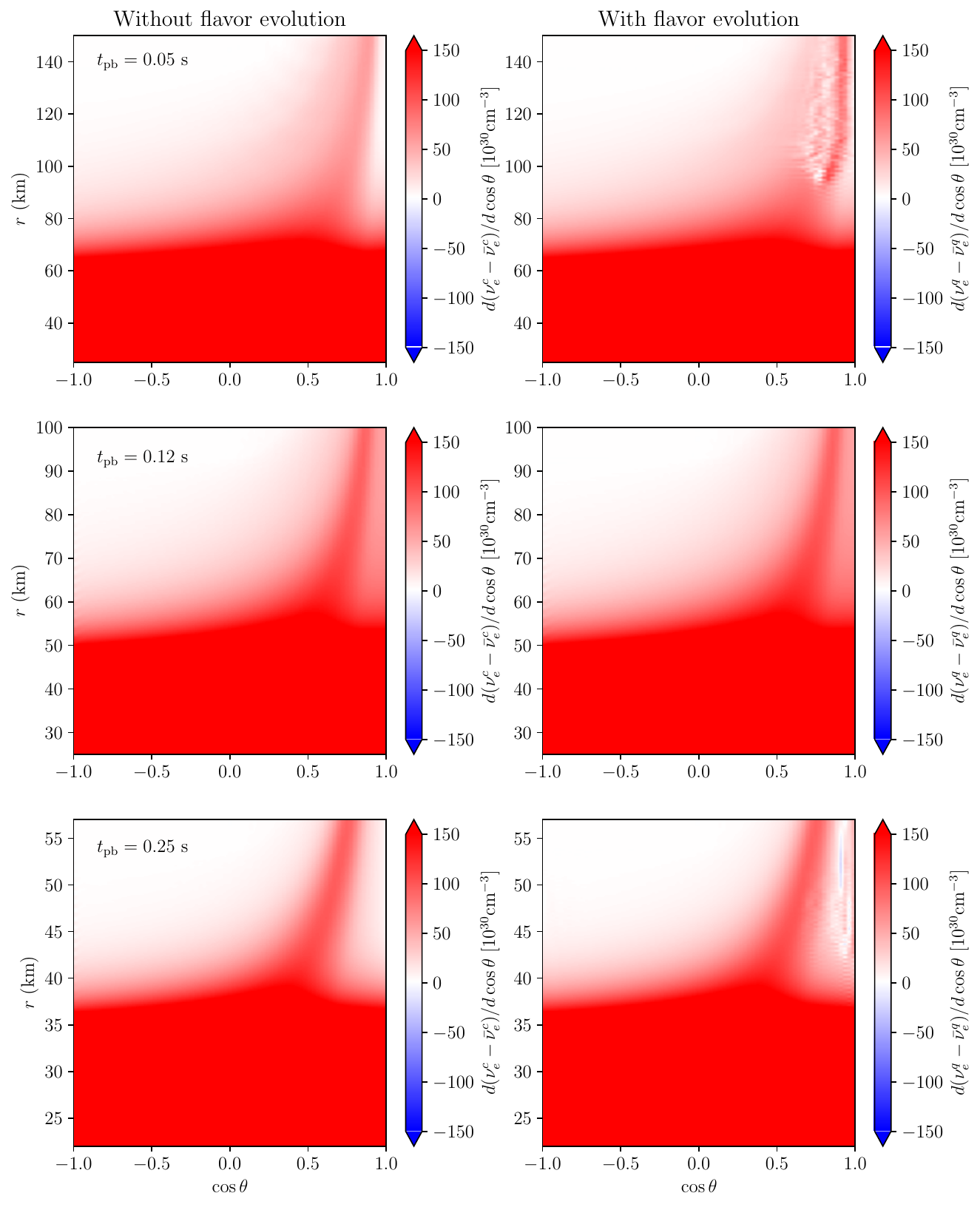}
\caption{{\it Left:}  Heatmaps of the energy-integrated ELN without flavor conversion ($\rho_{ee}-\bar\rho_{ee}$, classical solution) for $t_{\textrm{pb}}=0.05$, $0.12$, and $0.25$~s, from top to bottom respectively. At small radii, the neutrino angular distributions of all flavors are isotropic. In the proximity of neutrino decoupling, the angular distributions start becoming forward peaked, but no ELN crossing develops. The white shade in the top-left corners corresponds to a region with no neutrinos. {\it Right:} Same as the left column, but with neutrino flavor conversion included ($\rho_{ee}-\bar\rho_{ee}$, quantum solution); the QKEs have been evolved for $t=1.25 \times 10^{-4}$~s. The plots in this figure have been obtained by averaging the flavor content between $t=1.0 \times 10^{-4}$ and $t=1.25 \times 10^{-4}$~s. For $t_{\mathrm{pb}}=0.12$~s, no flavor evolution is seen, but for $t_{\mathrm{pb}}=0.05$ and $0.25$~s flavor evolution occurs due to vacuum mixing, despite the absence of ELN crossings.}
\label{Fig1a}
\end{figure}

\begin{figure}
\centering
\includegraphics[width=0.9\textwidth]{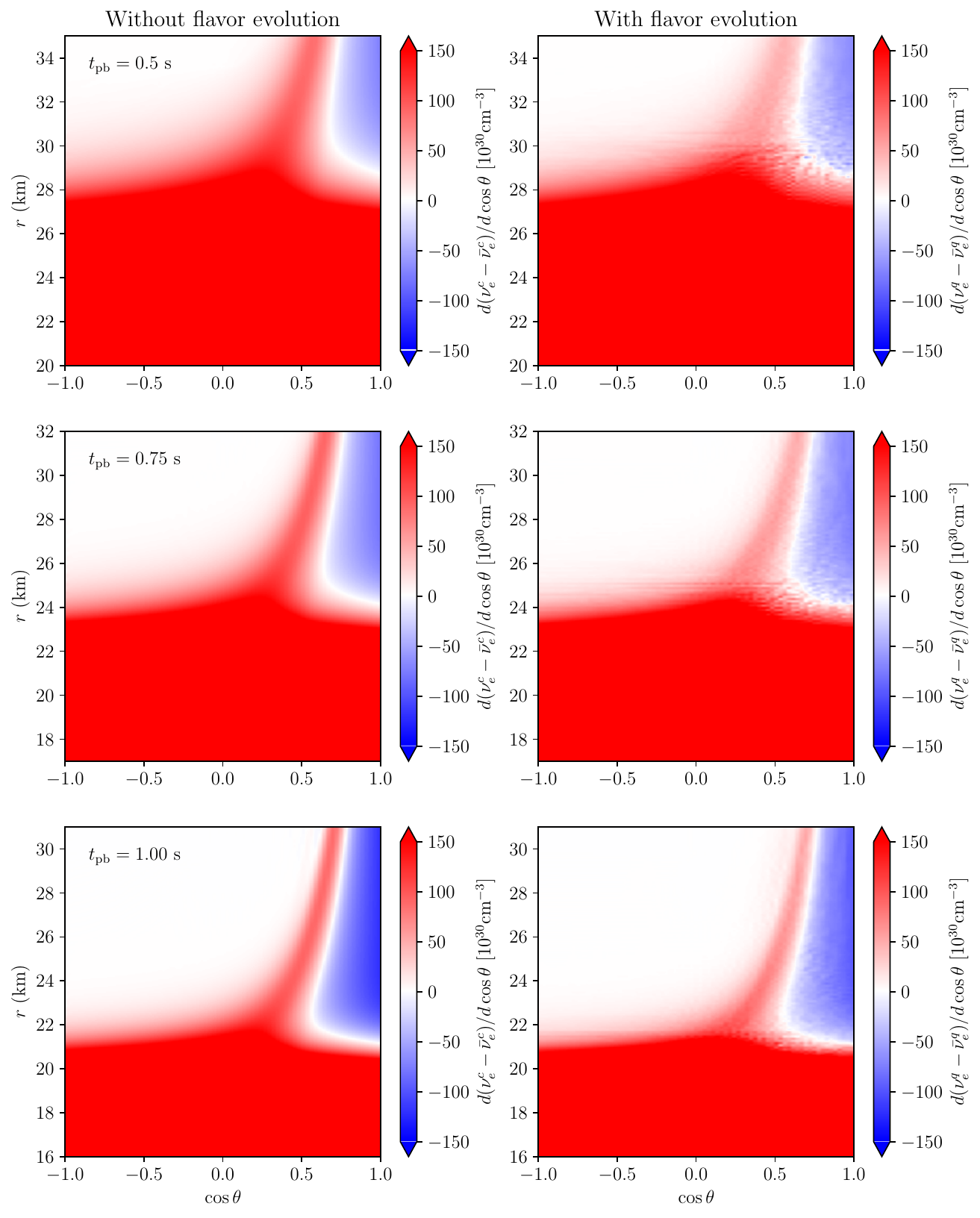}
\caption{Same as Fig.~\ref{Fig1a} but for $t_{\textrm{pb}}=0.5$, $0.75$, and $1$~s, from top to bottom respectively. In the proximity of neutrino decoupling, the angular distributions start becoming forward peaked and ELN crossings develop. The latter are marked by the white contour separating the red and blue regions. In the proximity of ELN crossings, fast flavor conversion develops as expected (the QKEs have been evolved for $t = 7.5 \times 10^{-5}$~s).  The plots have been obtained by averaging the flavor content between $t=5 \times 10^{-5}$ and $t=7.5 \times 10^{-5}$~s.
\label{Fig1b}
}
\end{figure}

The left panels of Figs.~\ref{Fig1a} and \ref{Fig1b} display the  heatmaps of the energy-integrated $\rho_{ee}-\bar{\rho}_{ee}$ in the plane spanned by $\cos\theta$ and $r$ in the absence of flavor conversion (i.e., the classical steady state configuration). One can see that, at small radii (before neutrino decoupling), the angular distributions of neutrinos and antineutrinos are isotropic with an overall excess of $\nu_e$. At larger radii, (anti)neutrinos start to decouple from matter entering the free streaming regime. As expected, because of the overall decrease of baryon density, neutrinos decouple at smaller radii as the post-bounce time increases. As (anti)neutrinos approach the free-streaming regime, an ELN crossing forms for $t_{\rm pb} \gtrsim 0.5$~s (visible from the white contour separating the red from the blue regions in the left panels of Fig.~\ref{Fig1b}). The presence of an ELN crossing is a necessary condition for flavor instability and becomes a sufficient condition, if periodic boundary conditions are assumed~\cite{Izaguirre:2016gsx, Morinaga:2018aug,Fiorillo:2024bzm}. 

\begin{figure}
\centering
\includegraphics[width=0.6\textwidth]{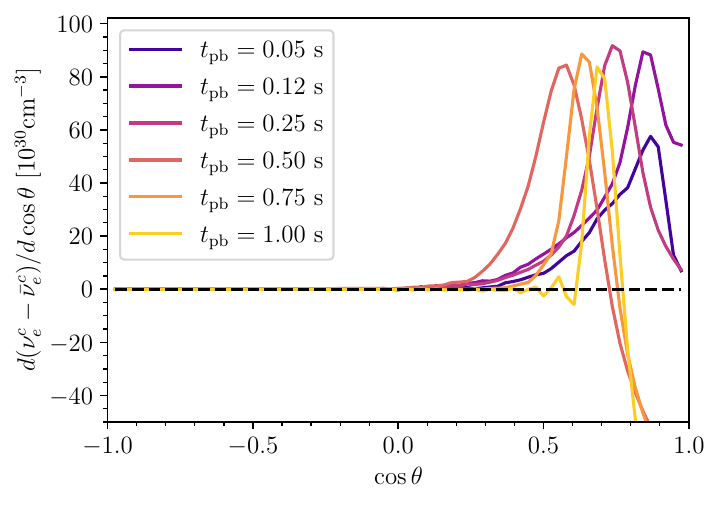}
\caption{Energy integrated ELN distributions as functions of $\cos\theta$ for our selected post-bounce time snapshots. An ELN crossing appears for $t_{\rm pb} \gtrsim 0.5$~s; the horizontal black dashed line marks ELN $=0$.}
\label{Fig0b}
\end{figure}

Figure~\ref{Fig0b} shows the energy-integrated ELN angular distributions, extracted at $r_{\rm max}$, as functions of $\cos\theta$ for our selection of post-bounce time snapshots. We can see that ELN crossings appear for $t_{\rm pb} \gtrsim 0.5$~s and are not present at early times. However, there is no monotonic trend as a function of radius for what concerns the ELN appearance at different $t_{\rm pb}$. It is worth noticing that Ref.~\cite{Shalgar:2023aca} found ELN crossings for all considered post-bounce time snapshots because it considered a single-energy solution of Eqs.~\ref{eom1} and \ref{eom2} for the sake of simplicity. We use this set of classical steady state configurations to investigate the differences in the flavor conversion outcome, as illustrated in the next section.

\section{Neutrino quantum kinetics for different  post-bounce times} \label{sec:results}
In this section, we present our results on the solution of the QKEs for different supernova time snapshots. First, we focus on the early accretion phase ($t_{\rm pb} \lesssim 0.5$~s) and then we investigate the quasi-steady state configuration during the late accretion phase ($t_{\rm pb} \gtrsim 0.5$~s). Finally, we artificially modify the collision term to investigate whether flavor equipartition is a generic outcome of flavor conversion.

\subsection{Early accretion phase: no ELN crossings, minimal flavor conversion}

\begin{figure}
 \includegraphics[width=0.49\textwidth]{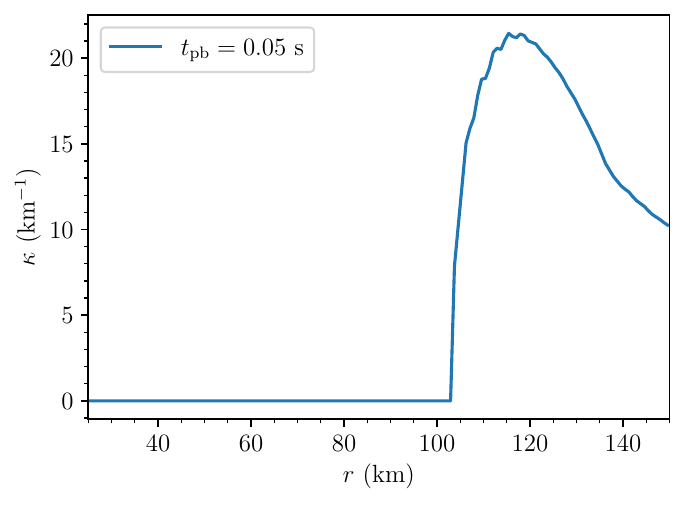}
 \includegraphics[width=0.49\textwidth]{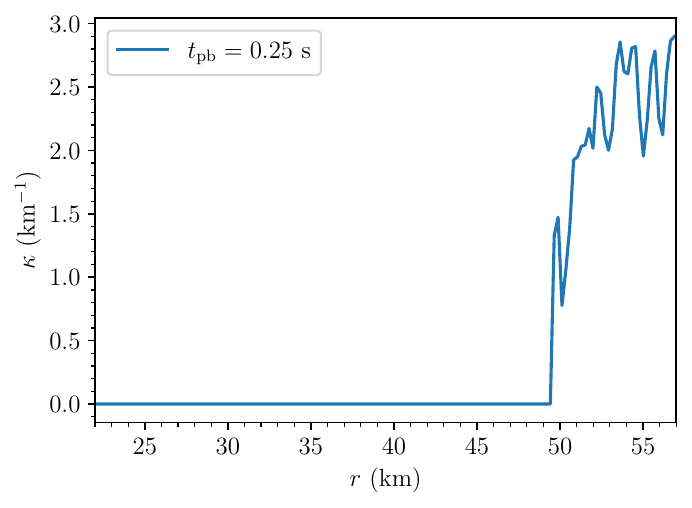}
 \caption{{\it Left:} Growth rate of the off-diagonal term of the density matrix for $t_{\mathrm{pb}} = 0.05$~s without rescaling of the self-interaction strength and adopting the average energy of the spectral energy distributions of $\nu_{e}$ and $\bar{\nu}_{e}$. The nonzero growth rate highlights the presence of a flavor instability, despite the absence of an ELN crossing (cf.~left panels of Fig.~\ref{Fig1a}). {\it Right:} Same as the left, but for $t_{\mathrm{pb}} = 0.25$~s. The growth rate is much smaller than that for $t_{\mathrm{pb}} = 0.05$~s, but it is non-zero. }
 \label{Figgrowth}
\end{figure}

Although we do not find ELN crossings at early times ($t_{\mathrm{pb}}\lesssim 0.5$~s), this does not imply the absence of flavor evolution as displayed in the right panels of Fig.~\ref{Fig1a} for $t_{\textrm{pb}}=0.05$ and $0.25$~s. 
Such flavor evolution is triggered by the vacuum mass term. Earlier work concluded that slow flavor evolution may be suppressed at early post-bounce times due to the large ratio of the self-interaction strength and the vacuum term~\cite{Esteban-Pretel:2008ovd,Sarikas:2012vb,Saviano:2012yh,Chakraborty:2011nf,Chakraborty:2011gd,Chakraborty:2014lsa}. However, as recently shown in Ref.~\cite{DedinNeto:2023ykt}, if the ELN distribution is on the verge of developing a crossing, the non-zero vacuum term and the neutrino-neutrino interaction term induce a collective instability. This is exactly what happens at $t_{\rm pb} = 0.05$ and $0.25$~s. We verify this conjecture by performing the linear stability analysis without rescaling the self-interaction strength. As shown in Fig.~\ref{Figgrowth}, a non-zero growth rate $k$ of the off-diagonal term of the density matrix is found.

The right panels of Fig.~\ref{Fig1a}  
show that, unlike for the flavor instability induced by an ELN crossing (see Sec.~\ref{sec:ELNevolution}),  flavor equipartition does not take place. 
This can be attributed to a combination of the ELN configurations and the fact that substantial flavor conversion cannot take place before (anti)neutrinos are advected out of the simulation shell.

\subsection{Late accretion phase: appearance of ELN crossings and fast flavor conversion}\label{sec:ELNevolution}

The right panels of Fig.~\ref{Fig1b} show the  heatmaps of the energy-integrated $\rho_{ee}-\bar\rho_{ee}$ at the simulation time $t=7.5 \times 10^{-5}$~s in the plane spanned by $\cos\theta$ and $r$ and in the presence of flavor conversion for $t_{\textrm{pb}}=0.5$, $0.75$ and  $1$~s (from top to bottom respectively). We find that flavor conversion develops in the proximity of the ELN crossings (the latter being denoted by the white region in between the red and blue ones) for $t_{\textrm{pb}}=0.5$, $0.75$ and $1$~s. 
Flavor conversion is therefore driven by fast flavor instabilities, as expected from the preliminary findings of Ref.~\cite{Shalgar:2023aca} in the single-energy approximation. In fact, Ref.~\cite{Shalgar:2023aca} found that the growth rate of the fast flavor instability is four orders of magnitude larger than the collisional instability one for these time snapshots (cf.~their Fig.~6). Reference~\cite{Shalgar:2023aca} also found that collisional instabilities are present within small spatial regions at smaller radii, yet we do not see any significant flavor conversion linked to them.

\begin{figure}
\centering
\includegraphics[width=0.9\textwidth]{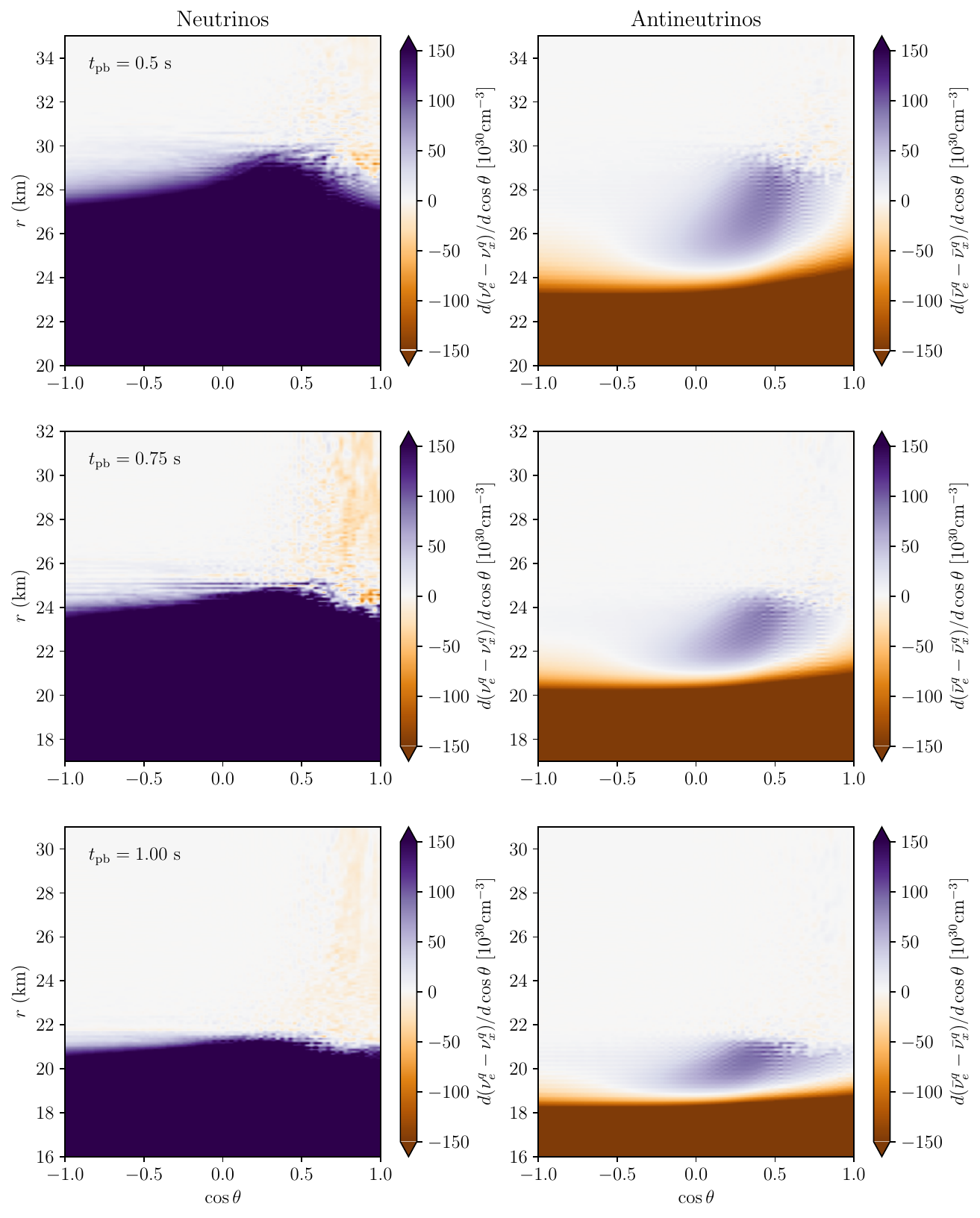}
 \caption{{\it Left:}  Heatmaps of the energy-integrated quasi-steady state configuration of $\rho_{ee}-\rho_{xx}$ (quantum solution) for $t_{\textrm{pb}}=0.5$, $0.75$, and $1$~s, from top to bottom respectively (the simulation has been evolved for $t=7.5\times 10^{-5}$~s) The plots have been obtained by averaging the flavor content between $t=5\times 10^{-5}$ and $t=7.5\times 10^{-5}$~s. {\it Right:} Same as the left column but for antineutrinos. As a result of flavor conversion, flavor equipartition is achieved at large radii, after neutrino decoupling, for neutrinos and antineutrinos ($\rho_{ee} \simeq \rho_{xx}$ and $\bar\rho_{ee} \simeq \bar\rho_{xx}$; see also Fig.~\ref{Fig4}).  }
\label{Fig2}
\end{figure}

\begin{figure}
\centering
\includegraphics[width=0.8\textwidth]{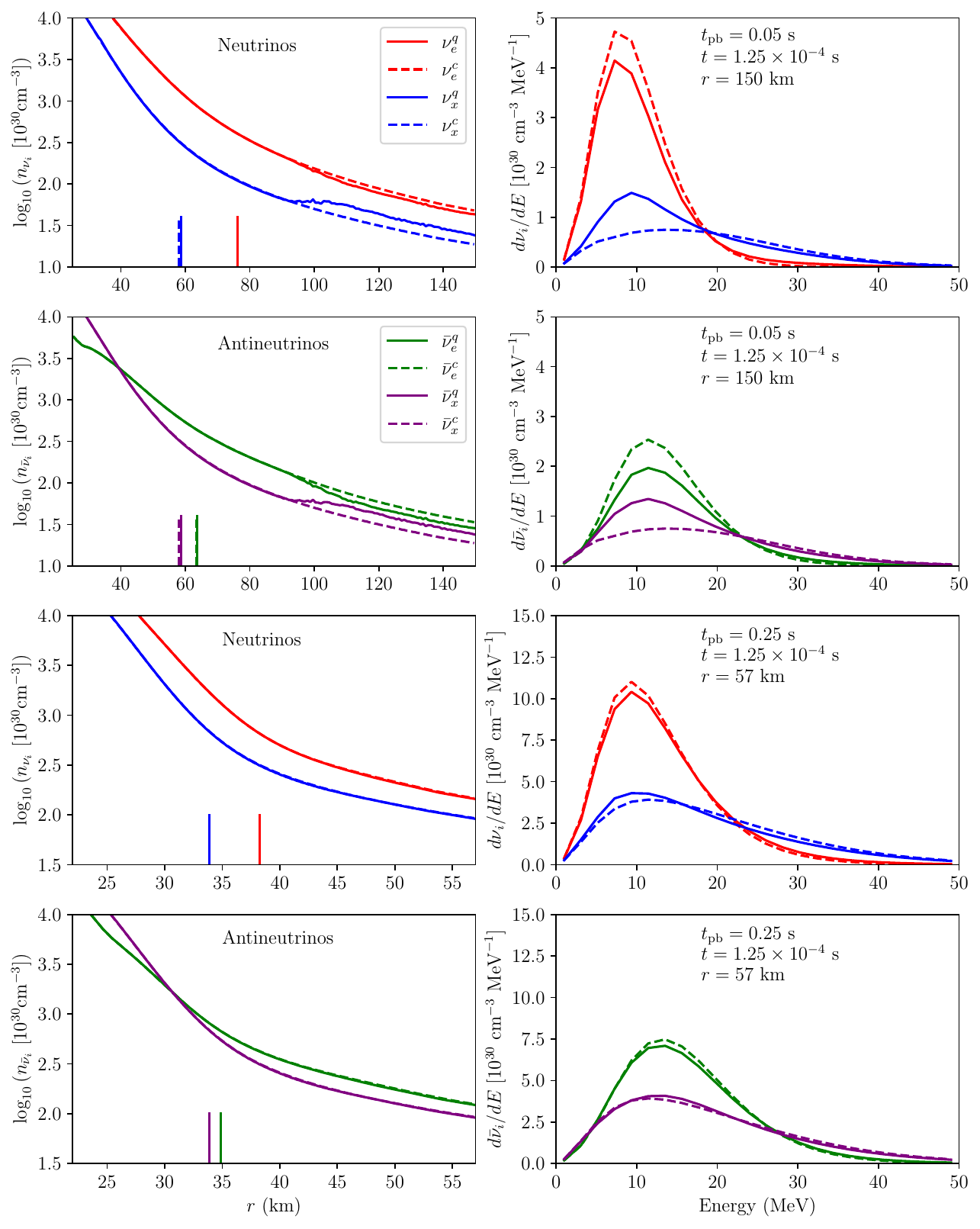}
		\caption{{\it Left: } Angle and energy integrated number densities for neutrinos (first and third row) and antineutrinos (second and fourth row) as functions of radius for  $t_{\rm pb} = 0.05$ (top panels) and  $0.25$~s (bottom panels). The dashed curves correspond to the classical steady state solution obtained in the absence of flavor conversion, while the solid curves represent the quasi-steady state configuration obtained including flavor conversion and computed after averaging between $t=1.0 \times 10^{-4}$ and $1.25 \times 10^{-4}$~s; during the quasi-steady state, we observe variations of the neutrino number densities with respect to their average that are smaller than $1\%$. 
  At early time $t_{\rm pb} \lesssim 0.5$ complete equipartition is not achieved. The radius of neutrino decoupling is marked through vertical lines in each plot and is minimally affected by flavor conversion. {\it Right:} Angle-integrated spectral energy distributions as functions of the neutrino energy, extracted at $r_{\textrm{max}}$ for the classical steady state solution (dashed curves) and the quasi-steady state one (solid curves). Due to the large difference between the electron neutrino and antineutrino emission properties and lack of ELN crossings, flavor equipartition is not reached.}
\label{Early8panel}
\end{figure}

\begin{figure}
\centering
\includegraphics[width=0.8\textwidth]{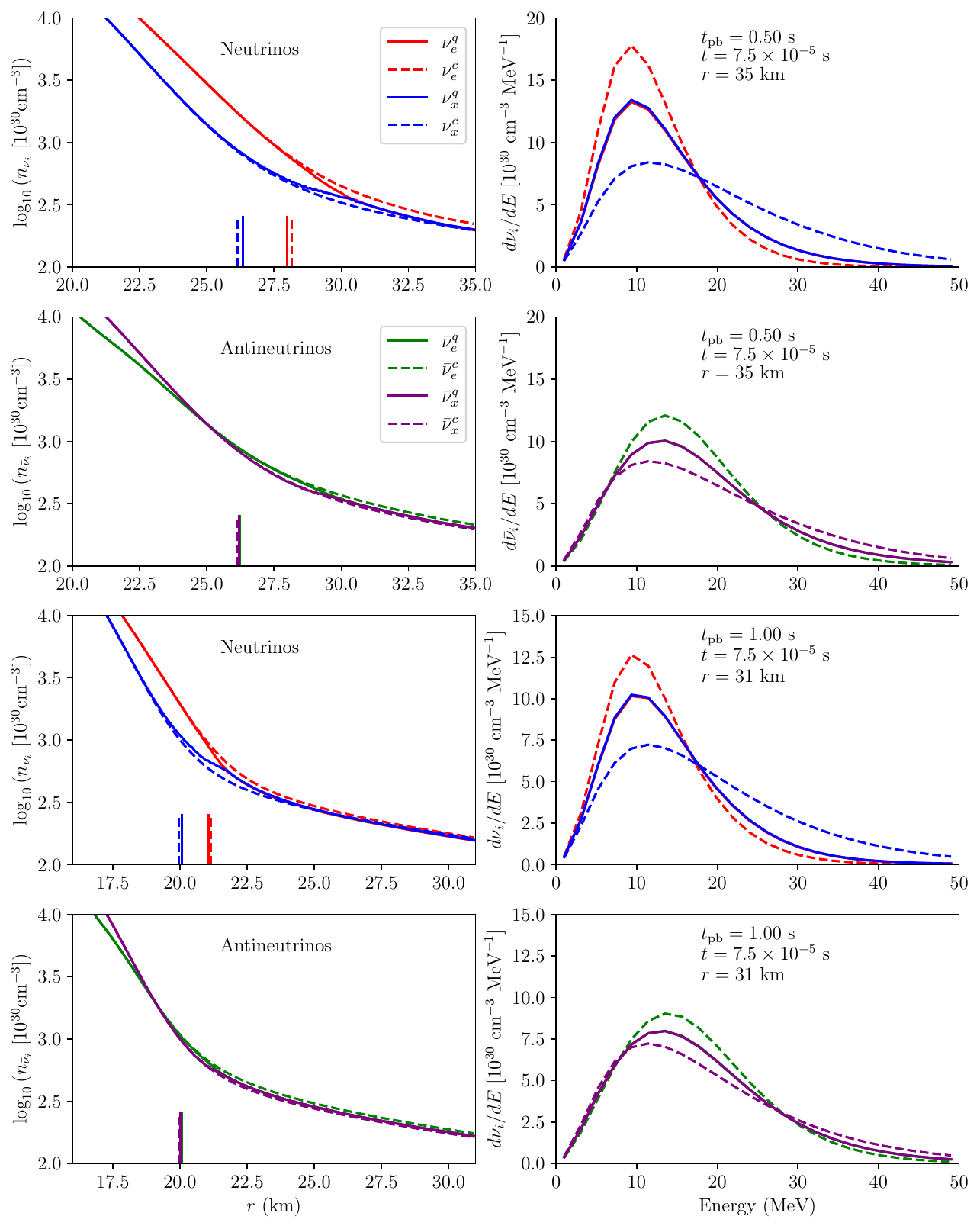}
\caption{{\it Left: } Angle and energy integrated number densities for neutrinos (first and third row) and antineutrinos (second and fourth row) as functions of radius for $t_{\rm pb} = 0.5$ (top panels) and $1$~s (bottom panels). The dashed curves correspond to the classical steady state solution obtained in the absence of flavor conversion, while the solid curves represent the quasi-steady state configuration obtained including flavor conversion by averaging between $t=5.0 \times 10^{-5}$ and  $t=7.5 \times 10^{-5}$~s; during the quasi-steady state, we observe variations of the neutrino number densities with respect to their average that are smaller than $1\%$. 
After neutrino decoupling flavor equipartition is achieved and the number densities of the different flavors become identical to each other. The radius of neutrino decoupling is marked through vertical lines in each plot and is minimally affected by flavor conversion. {\it Right:} Angle-integrated spectral energy distributions as functions of the neutrino energy, extracted at $r_{\textrm{max}}$ for the classical steady state solution (dashed curves) and the quasi-steady state one (solid curves). Because of flavor conversion, the spectral energy distribution of $\nu_{e}$ ($\bar{\nu}_{e}$) almost coincides with the one of $\nu_{x}$ ($\bar{\nu}_{x}$) for both post-bounce times.}
\label{Fig4}
\end{figure}
In order to investigate whether flavor equipartition is achieved as a result of flavor conversion~\footnote{ It should be noted that, in the presence of collisions, the lepton number is not conserved in the neutrino sector.}, Fig.~\ref{Fig2} displays the energy-integrated $\rho_{ee}-\rho_{xx}$ for the quasi-steady-state solution obtained after flavor conversion for neutrinos (antineutrinos) on the left (right) panels. Note that, in the region of neutrino trapping, there is an excess of $\nu_{e}$ over $\nu_{x}$ due to the positive chemical potential of $\nu_{e}$; while the opposite is true for antineutrinos. For all analyzed configurations for $t_{\rm pb} \gtrsim 0.5$~s, we find flavor equipartition (i.e.~the angular and spectral energy distributions of the $e$ and $x$ flavors approach each other, becoming comparable) is achieved for neutrinos and antineutrinos. This is also visible from the left panels of Fig.~\ref{Fig4}, where the radial evolution of the neutrino number density of the different flavors after flavor conversion (quantum solution) is compared to the one obtained neglecting flavor conversion (classical solution). In particular, as shown in the right panels of Fig.~\ref{Fig4} for $t_{\rm pb} = 0.5$ (top two panels) and $1$~s (bottom two panels), the angle-integrated energy distribution of $\nu_{e}$ ($\bar{\nu}_{e}$) is identical to the one of $\nu_{x}$ ($\bar{\nu}_{x}$) in the quasi-steady state configuration and at large radii ($r_{\textrm{max}}$ has been chosen to extract the spectral distributions plotted in Fig.~\ref{Fig4}, although any radius after neutrino decoupling shows a similar trend).

In the left panels of Fig.~\ref{Fig4}, the decoupling radii for each neutrino flavor are marked by vertical lines. The decoupling radius has been computed as the radius at which the flux factor is equal to $1/3$~\cite{Shalgar:2022rjj,Tamborra:2017ubu,Wu:2017drk}. We can see that the decoupling radius is minimally affected by flavor conversion (cf.~dashed vs.~solid lines); in fact, as shown in Ref.~\cite{Shalgar:2022rjj}, the impact of flavor conversion on the decoupling physics strongly depends on the initial ELN distribution and how similar the electron and non-electron flavor emission properties are in the absence of flavor conversion.

\subsection{Flavor equipartition is not a generic flavor outcome of fast flavor conversion}\label{sec:tweak}

\begin{figure}
\centering
\includegraphics[width=0.8\textwidth]{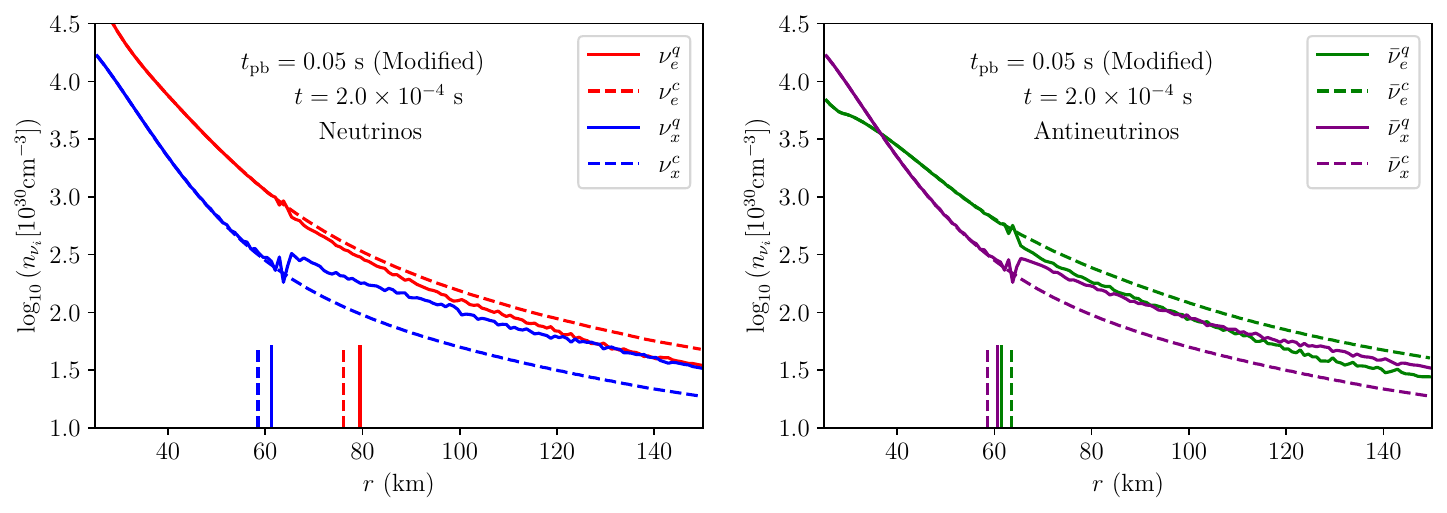}
\caption{{\it Left:} Angle and energy integrated neutrino number densities for each flavor as functions of radius for the modified configuration obtained by changing the emission term of $\bar{\nu}_{e}$ by $20\%$ at $t_{\textrm{pb}}=0.05$~s. The quasi-steady state is extracted at $t = 2 \times 10^{-4}$~s (simulation time), starting from the classical steady state solution. It can be seen that $\nu_{e}$ does not reach equipartition with $\nu_x$. {\it Right:} The same as the left panel, but for antineutrinos.
}
\label{Fig6}
\end{figure}

The neutrino properties obtained in the late accretion phase tend to be similar to each other for the different post-bounce times that we have considered (cf.~Figs.~\ref{Fig1a}, \ref{Fig1b} and \ref{Fig4}). In order to assess whether flavor equipartition should be expected to be the general flavor outcome of any quasi-steady state configuration, we 
increase the emission term of $\bar{\nu}_{e}$ at $t_{\textrm{pb}}=0.05$~s by $20\%$ to induce an ELN crossing. Solving the QKEs numerically, we find that flavor equipartition is not reached in the quasi-steady state solution, as shown in Fig.~\ref{Fig6}. 

This flavor outcome can be explained by looking at the number density of $\nu_{e}$ and $\bar{\nu}_{e}$ in the region of flavor instability: the two differ substantially compared to the configurations obtained at later post-bounce times for which flavor equipartition was found. 
The observed quasi-steady state configuration is due to the fact that the collision term tends to enhance the $\nu_e$ number density and suppress the $\nu_x$ one. On the other hand, flavor conversion tends to push $\nu_e$ and $\nu_x$ towards each other. As a result, the $\nu_x$ number density tends to change more (moving upwards) in the quasi-steady state configuration than the $\nu_e$ one with respect to the classical solution, being the impact of the collision term smaller for the non-electron flavors. As a consequence, the decoupling radii are also more largely affected by flavor conversion than observed for the other flavor configurations (cf.~Fig.~\ref{Fig1b}).

\begin{figure}
\includegraphics[width=0.49\textwidth]{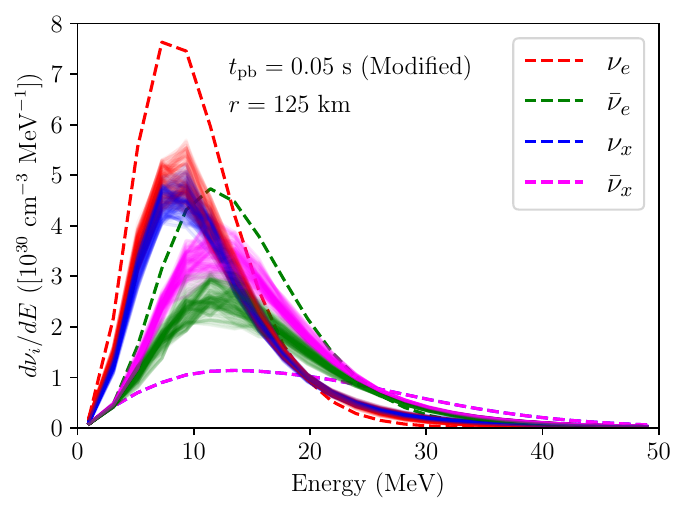}
\includegraphics[width=0.49\textwidth]{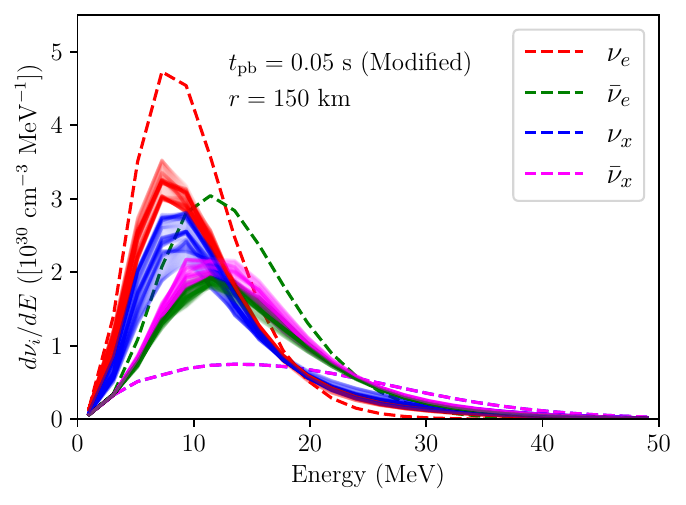}
\caption{{\it Left:} Angle-integrated spectral energy distributions plotted for $200$ selected simulation time steps between $1 \times 10^{-4}$ and $2 \times 10^{-4}$~s, and extracted at $r=125$~km for the modified configuration obtained by increasing by $20\%$ the emission term of $\bar{\nu}_{e}$ at $t_{\textrm{pb}}=0.05$~s. The energy spectra fluctuate in time by almost $10\%$, with the $\nu_{e}$ spectrum going approximately in equipartition with $\nu_{x}$. However, equipartition in the antineutrino sector is not achieved. {\it Right:} Same as the left panel, but for $r=150$~km. A comparable variation of $10\%$ in the energy distributions is observed in time once the quasi-steady state configuration is reached. In this case, the antineutrinos approach equipartition, but this is not the case for neutrinos.}
\label{Fig7}
\end{figure}

Figure~\ref{Fig7} shows the spectral energy distributions extracted for $200$ selected time steps at $r=125$~km and $150$~km, in order to investigate the temporal evolution of the quasi-steady flavor configuration. The neutrino spectra continue to oscillate in time, changing by as much as $10\%$. However, the system is approximately in partial equipartition for the neutrino sector and not for the antineutrino sector at $r=125$~km (cf.~left panel). This trend changes as a function of radius, e.g.~the opposite is true for $r=150$~km (cf.~right panel), and in general clearly shows that flavor equipartition is not achieved for this flavor configuration.

We note that the time over which flavor evolution occurs is much shorter than the time scale over which the collision term changes. The continued temporal evolution of the spectral energy distributions 
suggests that the physically relevant quantities with potential feedback on the source physics should be the average of the bands in Fig.~\ref{Fig7}.

\section{Discussion} \label{sec:discussion}
\begin{figure}
\includegraphics[width=0.49\textwidth]{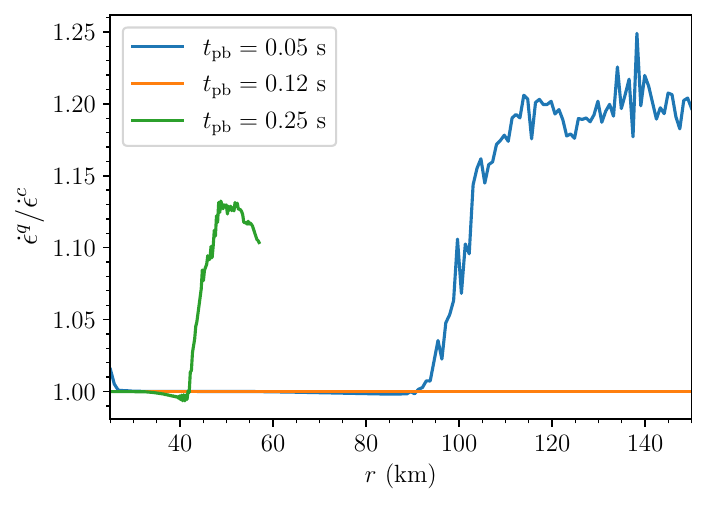}
\includegraphics[width=0.49\textwidth]{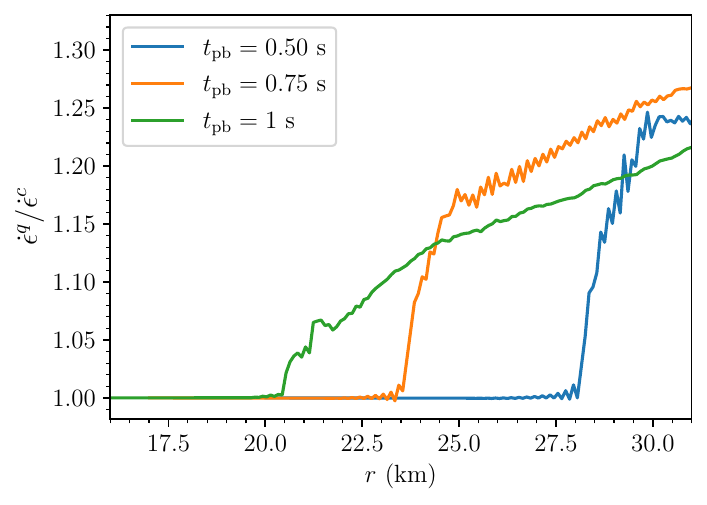}
\caption{{\it Left:} Ratio between the neutrino heating rates obtained adopting the quasi-steady state angle-integrated neutrino distributions after solving the QKEs and the ones computed from the classical solution (i.e., neglecting flavor conversion) for early times, $t_{\rm{pb}}= 0.05$, $0.12$ and $0.25$~s. At radii larger than the one of neutrino decoupling, the heating rate is increased by $15$--$25\%$, except for the case of $t_{\rm{pb}}= 0.12$~s. The heating ratios are obtained by averaging the flavor content that has been evolved between $t=1.0 \times 10^{-4}$ and $t=1.25\times 10^{-4}$~s. {\it Right:} Same as the left panel but for late accretion times, $t_{\rm{pb}}= 0.5$, $0.75$ and $1$~s (the flavor content used has been averaged between simulation times of $t=5.0 \times 10^{-5}$ and $t=7.5\times 10^{-5}$~s).
\label{Fig8}}
\end{figure}

In order to assess the possible impact of our findings on neutrino flavor conversion physics on the supernova mechanism, Fig.~\ref{Fig8} shows the ratio of the neutrino heating rates 
obtained relying on the quasi-steady state configuration when the QKEs are solved with and without the commutator term encapsulating the flavor conversion physics. The heating rate is defined as 
 \begin{eqnarray}
 \dot{\epsilon} &=& \dot{\epsilon}_{\nu_{e}} + \dot{\epsilon}_{\bar{\nu}_{e}}  \quad \textrm{where,} \\
 \dot{\epsilon}_{\nu_{e}} &=& \sigma_{0} \left(\frac{1+3g_{\textrm{A}}}{4}\right)\int_{0}^{\infty} dE \left(\frac{E+Q}{{m_{e}c^{2}}}\right)^{2} \sqrt{(E+Q)^{2}-m_{e}^{2}}\nonumber\\
 &\times& \left[1-\left(\frac{m_{e}c^{2}}{E+Q}\right)\right]^{\frac{1}{2}}
 \left(1-1.01\frac{E}{m_{n}}\right)\left(1-f_{e^{-}}\right)\frac{d n_{\nu_{e}}}{dE} \\
  \dot{\epsilon}_{\bar{\nu}_{e}}  & = & \sigma_{0} \left(\frac{1+3g_{\textrm{A}}}{4}\right)\int_{m_{e}+Q}^{\infty} dE \left(\frac{E-Q}{{m_{e}c^{2}}}\right)^{2} \sqrt{(E-Q)^{2}-m_{e}^{2}}\nonumber\\
 &\times&
 \left[1-\left(\frac{m_{e}c^{2}}{E-Q}\right)\right]^{\frac{1}{2}}
 \left(1-7.1\frac{E}{m_{p}}\right)\frac{d n_{\bar{\nu}_{e}}}{dE}\ ,
 \end{eqnarray}
where  $E$ denotes the energy or $\nu_{e}$ or $\bar{\nu}_{e}$, $Q = 1.2933$~MeV denotes the $Q$-value of the $\beta$-reaction, $m_{e}=0.511$~MeV is the mass of the electron, $\sigma_{0}$ is the characteristic neutrino interaction cross section ($4G_{\textrm{F}}^{2} m_{e}^{2}/\pi \approx 1.7 \times 10^{-44}$~cm$^{2}$), $(1-f_{e^{-}})$ is the electron Pauli-blocking factor, and $g_{A}=1.27$ is the axial coupling. 
This heating rate has been obtained by integrating the product of cross sections~\cite{thompsonthesis} and the energy released per interaction.

Figure~\ref{Fig8} shows that, for all cases except $t_{\textrm{pb}}=0.12$~s (for which we do not see any flavor conversion), an increase in the heating rate by $15$--$25\%$ occurs.
This is not surprising since the flavor evolution increases the number density of electron type neutrinos in the high energy tail of the spectral distribution, and a larger heating rate is expected for larger neutrino energy. 
While we do not compute the neutrino cooling rate and do not take into account the feedback that the neutrino flavor conversion physics could have on the supernova hydrodynamic background, our results suggest a non-trivial impact of neutrino flavor conversion physics on the supernova mechanism, supporting earlier parametric studies~\cite{Ehring:2023abs,Ehring:2023lcd,Nagakura:2023mhr}.

Our findings should be considered with caution as they have been obtained relying on a spherically symmetric core-collapse supernova model. While we observe a non-trivial appearance of ELN crossings as a function of the post-bounce time in the classical solution of the Boltzmann equations, multi-dimensional core-collapse supernova simulations show evidence of large scale asymmetries in the neutrino emission (e.g.~due to the LESA instability~\cite{Tamborra:2014aua}) and the location of flavor instabilities may be impacted~\cite{Akaho:2023brj}. 

In order to facilitate the numerical solution of the QKEs, we choose to attenuate the self-interaction strength by a constant factor $10^{-3}$ (see Eq.~\ref{eq:Hnunu}), nevertheless for all post-bounce times we have that $\mu \gg \omega$ at neutrino decoupling. When flavor equipartition is reached, it is unlikely that the rescaling of the self-interaction strength can affect the results. However, for the earlier accretion phase ($t_{\textrm{pb}} \lesssim 0.5$~s), flavor equipartition is not reached, and it is possible that this scaling might slightly affect the exact final flavor outcome and the time needed to reach a quasi-steady state solution.

The energy-dependent solution of the QKEs has consequences on the development of ELN crossings (cf.~the corresponding single-energy solution presented in Ref.~\cite{Shalgar:2023aca}) and therefore the presence of flavor instabilities; yet, for all post-bounce times, we find that a quasi-steady state configuration is achieved---cf.~also Refs.~\cite{Kato:2023dcw,Nagakura:2022kic}. These findings support recent attempts aiming at forecasting the 
quasi-steady state flavor configuration due to flavor conversion or designing methods to incorporate such physics into neutrino-radiation-hydrodynamic simulations~\cite{Zaizen:2023ihz,Zaizen:2022cik,Xiong:2024pue,Abbar:2024ynh,Nagakura:2022xwe,Nagakura:2023jfi,Abbar:2023ltx}; on the other hand, caution should be adopted in the forecast of the quasi-steady state neutrino configuration because of the dependence of the latter on the boundary conditions of the simulation box and features of the ELN distribution~\cite{Cornelius:2023eop,Zaizen:2023ihz,Shalgar:2022lvv}. 

We find that flavor equipartition is obtained for $t_{\rm{pb}}\gtrsim 0.5$~s. Such results may seem to confirm recent work in this direction~\cite{Xiong:2024tac,Xiong:2024pue,Martin:2021xyl,Zaizen:2023ihz,Zaizen:2022cik,Xiong:2023vcm,Grohs:2022fyq,Richers:2021xtf,Richers:2022bkd,Bhattacharyya:2020jpj,Bhattacharyya:2022eed,Wu:2021uvt}, obtained relying on local box simulations and finding near flavor equipartition on one side of the ELN crossing. However, as shown through the flavor configuration obtained by tweaking the collision term in Sec.~\ref{sec:tweak}, we stress that flavor equipartition depends on the initial conditions and should not be considered a general outcome of global flavor simulations~\cite{Cornelius:2023eop,Shalgar:2022rjj,Shalgar:2022lvv}. 

Similar to Ref.~\cite{Nagakura:2023xhc}, we explore the flavor conversion physics for a range of neutrino flavor configurations linked to different post-bounce times. While our results are in general agreement with the ones of Ref.~\cite{Nagakura:2023xhc}, we scan a broader range of post-bounce times highlighting time-dependent features of the flavor outcome. Moreover, differently from Ref.~\cite{Nagakura:2023xhc}, we do not artificially tweak the electron abundance to favor the appearance of ELN crossings; this has implications on the relative relevance between collisional and fast flavor instabilities. We also confirm the preliminary findings of Refs.~\cite{Shalgar:2023aca,Akaho:2023brj,Nagakura:2023xhc}, where it was concluded that collisional instabilities do not affect the flavor composition significantly.

\section{Conclusions} \label{sec:conclusions}
Understanding the neutrino flavor distribution resulting from flavor conversion occurring in the dense core of a supernova is crucial in order to investigate the impact of this physics on the supernova explosion mechanism. We numerically solve the neutrino quantum kinetic equations (QKEs) in a multi-energy and multi-angle framework and within a spherically symmetric shell, tracking neutrino decoupling.

We solve the QKEs adopting as inputs static hydrodynamic backgrounds extracted from a one-dimensional core-collapse supernova simulation with a mass of $18.6\ M_\odot$ at post-bounce times $t_{\rm{pb}} = 0.05$, $0.12$, $0.25$, $0.5$, $0.75$, and $1$~s. First, neglecting flavor conversion, we compute the angular distributions of neutrinos and antineutrinos, following Ref.~\cite{Shalgar:2023aca}. Differently from Ref.~\cite{Shalgar:2023aca} and because of the multi-energy numerical solution of the QKEs, we find that crossings in the electron neutrino lepton number (ELN) angular distributions appear for $t_{\rm{pb}} \gtrsim 0.5$~s, but not at earlier post-bounce times. 

We adopt the classical steady-state solution to solve the QKEs, this time taking into account neutrino flavor conversion; however, for the sake of simplicity, we neglect the impact of the matter (non-neutrino) background on the flavor evolution. Fast flavor conversion is responsible for making the angular and spectral energy distribution of $\nu_e$ ($\bar\nu_e$) equal to the one of $\nu_{x}$ ($\bar\nu_x$) for $t_{\rm{pb}} \gtrsim 0.5$~s, i.e.~flavor equipartition is reached.
Despite the changes in the neutrino energy and angular distributions of all flavors, we find that the flavor-dependent neutrino decoupling radii are negligibly affected for the investigated flavor configurations; this is because of the relatively similar properties between the electron and non-electron flavors characterizing the late accretion phase that facilitate flavor equipartition, but overall negligibly affect the energy and angle integrated flavor-dependent number density at decoupling. On the other hand, ELN crossings are not found for $t_{\rm{pb}} \lesssim 0.5$~s, but flavor conversion starts because of slow collective instabilities. 

Remarkably, flavor equipartition between the angular and/or energy distributions of the electron and non-electron flavors is not a generic outcome of fast flavor conversion, but rather the result of the interplay between collisions, flavor conversion, and advection, when the neutrino emission properties of the electron and non-electron flavors are reasonably comparable to each other. Modifying the collision term artificially for a time snapshot extracted at $t_{\rm{pb}} \lesssim 0.5$~s to induce a crossing in the ELN angular distribution, we find that flavor equipartition is not achieved. This is because the differences in the emission properties of different flavors are larger in the early accretion phase and the interplay between flavor conversion and collisions cannot equilibrate the electron and non-electron flavors. 

Our findings should be expanded to draw conclusions on the flavor conversion phenomenology expected in a broader range of core-collapse supernova models. Moreover, a self-consistent implementation of flavor conversion in supernova hydrodynamic simulations may dynamically change the neutrino emission properties as a function of the post-bounce time, in turn possibly affecting the flavor conversion phenomenology. Despite its intrinsic limitations, our work highlights a very interesting, non-trivial flavor phenomenology in core-collapse supernovae, calling for further investigation in order to assess its impact on the neutrino-driven explosion mechanism and nucleosynthesis.

\acknowledgments
We thank Marie Cornelius, Damiano Fiorillo, Georg Raffelt, Guenther Sigl, Meng-Ru Wu, and Zewei Xiong for  their comments on the manuscript.  This project has received support from the Danmarks Frie Forskningsfond (Project 
No.~8049-00038B), the European Union (ERC, ANET, Project No.~101087058), and the Deutsche Forschungsgemeinschaft through Sonderforschungbereich SFB 1258 ``Neutrinos and Dark Matter in Astro- and Particle Physics'' (NDM). 
Views and opinions expressed are those of the authors only and do not necessarily reflect those of the European Union or the European Research Council. Neither the European Union nor the granting authority can be held responsible for them. The Tycho supercomputer hosted at the SCIENCE HPC Center at the University of Copenhagen was used for supporting the numerical simulations presented in this work.

\appendix
\section{Comparison of the neutrino  number densities with the output of the supernova model}
\label{comparison}

\begin{figure}
\includegraphics[width=0.99\textwidth]{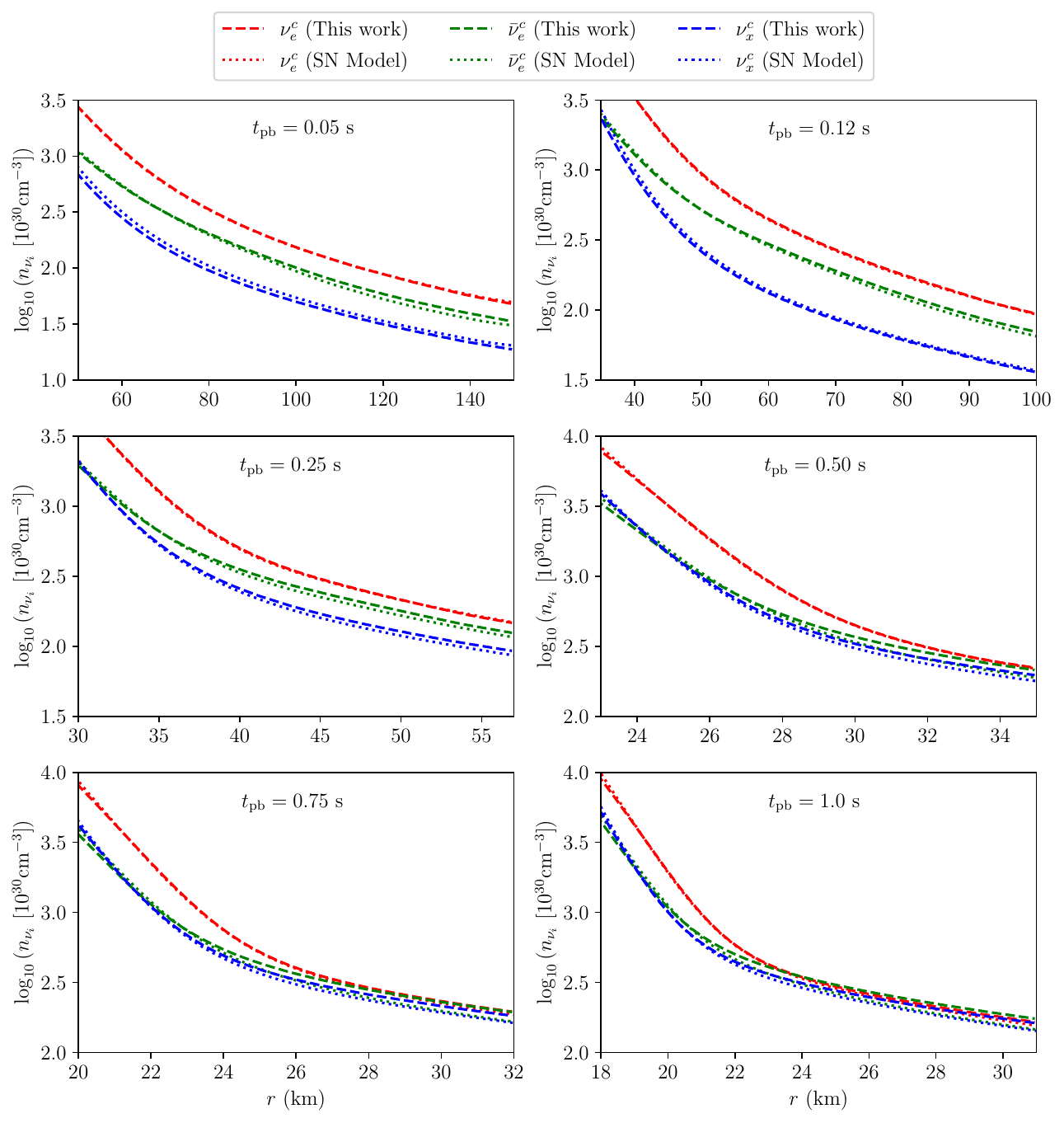}
\caption{Radial evolution of the  angle and energy integrated neutrino number densities computed in this work without neutrino mixing (dashed lines) and the ones extracted from the supernova simulation  (dotted). The red, green, and blue lines represent the number densities of $\nu_{e}$, $\bar{\nu}_{e}$, and $\nu_{x}$, respectively. 
Our solution of the QKEs, in the absence of flavor conversion, allows us to reproduce the local neutrino densities that are in extremely good agreement with the outputs of the supernova simulation. In some cases, the dotted line is not visible as it falls exactly behind the dashed one.}
\label{Figcomp}
\end{figure}

In this appendix, we provide a comparison between the number densities resulting from our benchmark hydrodynamical supernova model~\cite{SNarchive} and the  neutrino number densities obtained by solving Eqs.~\ref{eom1} and \ref{eom2}  with $H=0$ and $\bar{H}=0$. Figure~\ref{Figcomp} shows the angle and energy integrated neutrino number densities for our  six selected post-bounce time steps  ($t_{\mathrm{pb}} = 0.05, 0.12,0.25,0.5,0.75,$ and $1.0$~s). The dashed lines show the number densities obtained in this work, whereas the dotted lines show the number densities from the hydrodynamical supernova model~\cite{SNarchive}. For all times and radii, the agreement is excellent with a difference between the two cases smaller than  $10\%$. We refer the reader to  Sec.~III and Appendix B of Ref.~\cite{Shalgar:2023aca} for more details and an analogous implementation within a single-energy approximation.

We stress, however, that the scope of this work is not to reproduce the outputs of the hydrodynamical simulations with a full Boltzmann approach. We aim to compute the neutrino angular distributions from first principles and explore the variations in the flavor conversion physics.

\bibliographystyle{JHEP}
\bibliography{multi-energy.bib}

\providecommand{\href}[2]{#2}\begingroup\raggedright\begin{thebibliography}{10}

\bibitem{Pantaleone:1994ns}
J.~T. Pantaleone, \emph{{Neutrino flavor evolution near a supernova's core}},
  \href{https://doi.org/10.1016/0370-2693(94)01369-N}{\emph{Phys. Lett. B}
  {\bfseries 342} (1995) 250--256},
  [\href{https://arxiv.org/abs/astro-ph/9405008}{{\ttfamily
  astro-ph/9405008}}].

\bibitem{Langacker:1992xk}
P.~Langacker and J.~Liu, \emph{{Standard Model contributions to the neutrino
  index of refraction in the early universe}},
  \href{https://doi.org/10.1103/PhysRevD.46.4140}{\emph{Phys. Rev. D}
  {\bfseries 46} (1992) 4140--4160},
  [\href{https://arxiv.org/abs/hep-ph/9206209}{{\ttfamily hep-ph/9206209}}].

\bibitem{Mirizzi:2015eza}
A.~Mirizzi, I.~Tamborra, H.-T. Janka, N.~Saviano, K.~Scholberg, R.~Bollig
  et~al., \emph{{Supernova Neutrinos: Production, Oscillations and Detection}},
  \href{https://doi.org/10.1393/ncr/i2016-10120-8}{\emph{Riv. Nuovo Cim.}
  {\bfseries 39} (2016) 1--112},
  [\href{https://arxiv.org/abs/1508.00785}{{\ttfamily 1508.00785}}].

\bibitem{Duan:2010bg}
H.~Duan, G.~M. Fuller and Y.-Z. Qian, \emph{{Collective Neutrino
  Oscillations}},
  \href{https://doi.org/10.1146/annurev.nucl.012809.104524}{\emph{Ann. Rev.
  Nucl. Part. Sci.} {\bfseries 60} (2010) 569--594},
  [\href{https://arxiv.org/abs/1001.2799}{{\ttfamily 1001.2799}}].

\bibitem{Chakraborty:2016yeg}
S.~Chakraborty, R.~Hansen, I.~Izaguirre and G.~G. Raffelt, \emph{{Collective
  neutrino flavor conversion: Recent developments}},
  \href{https://doi.org/10.1016/j.nuclphysb.2016.02.012}{\emph{Nucl. Phys. B}
  {\bfseries 908} (2016) 366--381},
  [\href{https://arxiv.org/abs/1602.02766}{{\ttfamily 1602.02766}}].

\bibitem{Tamborra:2020cul}
I.~Tamborra and S.~Shalgar, \emph{{New Developments in Flavor Evolution of a
  Dense Neutrino Gas}},
  \href{https://doi.org/10.1146/annurev-nucl-102920-050505}{\emph{Ann. Rev.
  Nucl. Part. Sci.} {\bfseries 71} (2021) 165--188},
  [\href{https://arxiv.org/abs/2011.01948}{{\ttfamily 2011.01948}}].

\bibitem{Richers:2022zug}
S.~Richers and M.~Sen, \emph{Fast Flavor Transformations}, pp.~1--17.
\newblock Springer Nature Singapore, Singapore, 2022.
\newblock \href{https://arxiv.org/abs/[2207.03561]}{{\ttfamily [2207.03561]}}.
\newblock 10.1007/978-981-15-8818-1\_125-1.

\bibitem{Duan:2005cp}
H.~Duan, G.~M. Fuller and Y.-Z. Qian, \emph{{Collective neutrino flavor
  transformation in supernovae}},
  \href{https://doi.org/10.1103/PhysRevD.74.123004}{\emph{Phys. Rev. D}
  {\bfseries 74} (2006) 123004},
  [\href{https://arxiv.org/abs/astro-ph/0511275}{{\ttfamily
  astro-ph/0511275}}].

\bibitem{Duan:2006an}
H.~Duan, G.~M. Fuller, J.~Carlson and Y.-Z. Qian, \emph{{Simulation of Coherent
  Non-Linear Neutrino Flavor Transformation in the Supernova Environment. 1.
  Correlated Neutrino Trajectories}},
  \href{https://doi.org/10.1103/PhysRevD.74.105014}{\emph{Phys. Rev. D}
  {\bfseries 74} (2006) 105014},
  [\href{https://arxiv.org/abs/astro-ph/0606616}{{\ttfamily
  astro-ph/0606616}}].

\bibitem{Sawyer:2005jk}
R.~F. Sawyer, \emph{{Speed-up of neutrino transformations in a supernova
  environment}}, \href{https://doi.org/10.1103/PhysRevD.72.045003}{\emph{Phys.
  Rev. D} {\bfseries 72} (2005) 045003},
  [\href{https://arxiv.org/abs/hep-ph/0503013}{{\ttfamily hep-ph/0503013}}].

\bibitem{Sawyer:2008zs}
R.~F. Sawyer, \emph{{The multi-angle instability in dense neutrino systems}},
  \href{https://doi.org/10.1103/PhysRevD.79.105003}{\emph{Phys. Rev. D}
  {\bfseries 79} (2009) 105003},
  [\href{https://arxiv.org/abs/0803.4319}{{\ttfamily 0803.4319}}].

\bibitem{Sawyer:2015dsa}
R.~F. Sawyer, \emph{{Neutrino cloud instabilities just above the neutrino
  sphere of a supernova}},
  \href{https://doi.org/10.1103/PhysRevLett.116.081101}{\emph{Phys. Rev. Lett.}
  {\bfseries 116} (2016) 081101},
  [\href{https://arxiv.org/abs/1509.03323}{{\ttfamily 1509.03323}}].

\bibitem{Chakraborty:2016lct}
S.~Chakraborty, R.~S. Hansen, I.~Izaguirre and G.~G. Raffelt,
  \emph{{Self-induced neutrino flavor conversion without flavor mixing}},
  \href{https://doi.org/10.1088/1475-7516/2016/03/042}{\emph{JCAP} {\bfseries
  03} (2016) 042}, [\href{https://arxiv.org/abs/1602.00698}{{\ttfamily
  1602.00698}}].

\bibitem{Izaguirre:2016gsx}
I.~Izaguirre, G.~G. Raffelt and I.~Tamborra, \emph{{Fast Pairwise Conversion of
  Supernova Neutrinos: A Dispersion-Relation Approach}},
  \href{https://doi.org/10.1103/PhysRevLett.118.021101}{\emph{Phys. Rev. Lett.}
  {\bfseries 118} (2017) 021101},
  [\href{https://arxiv.org/abs/1610.01612}{{\ttfamily 1610.01612}}].

\bibitem{Johns:2021qby}
L.~Johns, \emph{{Collisional Flavor Instabilities of Supernova Neutrinos}},
  \href{https://doi.org/10.1103/PhysRevLett.130.191001}{\emph{Phys. Rev. Lett.}
  {\bfseries 130} (2023) 191001},
  [\href{https://arxiv.org/abs/2104.11369}{{\ttfamily 2104.11369}}].

\bibitem{Morinaga:2021vmc}
T.~Morinaga, \emph{{Fast neutrino flavor instability and neutrino flavor lepton
  number crossings}},
  \href{https://doi.org/10.1103/PhysRevD.105.L101301}{\emph{Phys. Rev. D}
  {\bfseries 105} (2022) L101301},
  [\href{https://arxiv.org/abs/2103.15267}{{\ttfamily 2103.15267}}].

\bibitem{Fiorillo:2024bzm}
D.~F.~G. Fiorillo and G.~G. Raffelt, \emph{{Theory of neutrino fast flavor
  evolution. I. Linear response theory and stability conditions}},
  \href{https://arxiv.org/abs/2406.06708}{{\ttfamily 2406.06708}}.

\bibitem{Hannestad:2006nj}
S.~Hannestad, G.~G. Raffelt, G.~Sigl and Y.~Y.~Y. Wong, \emph{{Self-induced
  conversion in dense neutrino gases: Pendulum in flavour space}},
  \href{https://doi.org/10.1103/PhysRevD.74.105010}{\emph{Phys. Rev. D}
  {\bfseries 74} (2006) 105010},
  [\href{https://arxiv.org/abs/astro-ph/0608695}{{\ttfamily
  astro-ph/0608695}}].

\bibitem{Fiorillo:2023mze}
D.~F.~G. Fiorillo and G.~G. Raffelt, \emph{{Slow and fast collective neutrino
  oscillations: Invariants and reciprocity}},
  \href{https://doi.org/10.1103/PhysRevD.107.043024}{\emph{Phys. Rev. D}
  {\bfseries 107} (2023) 043024},
  [\href{https://arxiv.org/abs/2301.09650}{{\ttfamily 2301.09650}}].

\bibitem{Johns:2022yqy}
L.~Johns and Z.~Xiong, \emph{{Collisional instabilities of neutrinos and their
  interplay with fast flavor conversion in compact objects}},
  \href{https://doi.org/10.1103/PhysRevD.106.103029}{\emph{Phys. Rev. D}
  {\bfseries 106} (2022) 103029},
  [\href{https://arxiv.org/abs/2208.11059}{{\ttfamily 2208.11059}}].

\bibitem{Padilla-Gay:2022wck}
I.~Padilla-Gay, I.~Tamborra and G.~G. Raffelt, \emph{{Neutrino fast flavor
  pendulum. II. Collisional damping}},
  \href{https://doi.org/10.1103/PhysRevD.106.103031}{\emph{Phys. Rev. D}
  {\bfseries 106} (2022) 103031},
  [\href{https://arxiv.org/abs/2209.11235}{{\ttfamily 2209.11235}}].

\bibitem{Fiorillo:2023ajs}
D.~F.~G. Fiorillo, I.~Padilla-Gay and G.~G. Raffelt, \emph{{Collisions and
  collective flavor conversion: Integrating out the fast dynamics}},
  \href{https://doi.org/10.1103/PhysRevD.109.063021}{\emph{Phys. Rev. D}
  {\bfseries 109} (2024) 063021},
  [\href{https://arxiv.org/abs/2312.07612}{{\ttfamily 2312.07612}}].

\bibitem{Lin:2022dek}
Y.-C. Lin and H.~Duan, \emph{{Collision-induced flavor instability in dense
  neutrino gases with energy-dependent scattering}},
  \href{https://doi.org/10.1103/PhysRevD.107.083034}{\emph{Phys. Rev. D}
  {\bfseries 107} (2023) 083034},
  [\href{https://arxiv.org/abs/2210.09218}{{\ttfamily 2210.09218}}].

\bibitem{Xiong:2022vsy}
Z.~Xiong, M.-R. Wu, G.~Mart\'\i{}nez-Pinedo, T.~Fischer, M.~George, C.-Y. Lin
  et~al., \emph{{Evolution of collisional neutrino flavor instabilities in
  spherically symmetric supernova models}},
  \href{https://doi.org/10.1103/PhysRevD.107.083016}{\emph{Phys. Rev. D}
  {\bfseries 107} (2023) 083016},
  [\href{https://arxiv.org/abs/2210.08254}{{\ttfamily 2210.08254}}].

\bibitem{Nagakura:2023xhc}
H.~Nagakura and M.~Zaizen, \emph{{Basic characteristics of neutrino flavor
  conversions in the postshock regions of core-collapse supernova}},
  \href{https://doi.org/10.1103/PhysRevD.108.123003}{\emph{Phys. Rev. D}
  {\bfseries 108} (2023) 123003},
  [\href{https://arxiv.org/abs/2308.14800}{{\ttfamily 2308.14800}}].

\bibitem{Shalgar:2023aca}
S.~Shalgar and I.~Tamborra, \emph{{Do neutrinos become flavor unstable due to
  collisions with matter in the supernova decoupling region?}},
  \href{https://doi.org/10.1103/PhysRevD.109.103011}{\emph{Phys. Rev. D}
  {\bfseries 109} (2024) 103011},
  [\href{https://arxiv.org/abs/2307.10366}{{\ttfamily 2307.10366}}].

\bibitem{Colgate:1966ax}
S.~A. Colgate and R.~H. White, \emph{{The Hydrodynamic Behavior of Supernovae
  Explosions}}, \href{https://doi.org/10.1086/148549}{\emph{Astrophys. J.}
  {\bfseries 143} (1966) 626}.

\bibitem{Bethe:1985sox}
H.~A. Bethe and J.~R. Wilson, \emph{{Revival of a stalled supernova shock by
  neutrino heating}}, \href{https://doi.org/10.1086/163343}{\emph{Astrophys.
  J.} {\bfseries 295} (1985) 14--23}.

\bibitem{Janka:2016fox}
H.~T. Janka, T.~Melson and A.~Summa, \emph{{Physics of Core-Collapse Supernovae
  in Three Dimensions: a Sneak Preview}},
  \href{https://doi.org/10.1146/annurev-nucl-102115-044747}{\emph{Ann. Rev.
  Nucl. Part. Sci.} {\bfseries 66} (2016) 341--375},
  [\href{https://arxiv.org/abs/1602.05576}{{\ttfamily 1602.05576}}].

\bibitem{Muller:2020ard}
B.~M\"uller, \emph{{Hydrodynamics of core-collapse supernovae and their
  progenitors}},
  \href{https://doi.org/10.1007/s41115-020-0008-5}{\emph{Astrophysics}
  {\bfseries 6} (2020) 3}, [\href{https://arxiv.org/abs/2006.05083}{{\ttfamily
  2006.05083}}].

\bibitem{Mezzacappa:2020oyq}
A.~Mezzacappa, E.~Endeve, O.~E. Bronson~Messer and S.~W. Bruenn,
  \emph{{Physical, numerical, and computational challenges of modeling neutrino
  transport in core-collapse supernovae}},
  \href{https://doi.org/10.1007/s41115-020-00010-8}{\emph{Liv. Rev. Comput.
  Astrophys.} {\bfseries 6} (2020) 4},
  [\href{https://arxiv.org/abs/2010.09013}{{\ttfamily 2010.09013}}].

\bibitem{Burrows:2020qrp}
A.~Burrows and D.~Vartanyan, \emph{{Core-Collapse Supernova Explosion Theory}},
  \href{https://doi.org/10.1038/s41586-020-03059-w}{\emph{Nature} {\bfseries
  589} (2021) 29--39}, [\href{https://arxiv.org/abs/2009.14157}{{\ttfamily
  2009.14157}}].

\bibitem{Ehring:2023lcd}
J.~Ehring, S.~Abbar, H.-T. Janka, G.~G. Raffelt and I.~Tamborra, \emph{{Fast
  neutrino flavor conversion in core-collapse supernovae: A parametric study in
  1D models}}, \href{https://doi.org/10.1103/PhysRevD.107.103034}{\emph{Phys.
  Rev. D} {\bfseries 107} (2023) 103034},
  [\href{https://arxiv.org/abs/2301.11938}{{\ttfamily 2301.11938}}].

\bibitem{Ehring:2023abs}
J.~Ehring, S.~Abbar, H.-T. Janka, G.~G. Raffelt and I.~Tamborra, \emph{{Fast
  Neutrino Flavor Conversions Can Help and Hinder Neutrino-Driven Explosions}},
  \href{https://doi.org/10.1103/PhysRevLett.131.061401}{\emph{Phys. Rev. Lett.}
  {\bfseries 131} (2023) 061401},
  [\href{https://arxiv.org/abs/2305.11207}{{\ttfamily 2305.11207}}].

\bibitem{Nagakura:2023mhr}
H.~Nagakura, \emph{{Roles of Fast Neutrino-Flavor Conversion on the
  Neutrino-Heating Mechanism of Core-Collapse Supernova}},
  \href{https://doi.org/10.1103/PhysRevLett.130.211401}{\emph{Phys. Rev. Lett.}
  {\bfseries 130} (2023) 211401},
  [\href{https://arxiv.org/abs/2301.10785}{{\ttfamily 2301.10785}}].

\bibitem{Johns:2023jjt}
L.~Johns, \emph{{Thermodynamics of oscillating neutrinos}},
  \href{https://arxiv.org/abs/2306.14982}{{\ttfamily 2306.14982}}.

\bibitem{Johns:2024dbe}
L.~Johns, \emph{{Subgrid modeling of neutrino oscillations in astrophysics}},
  \href{https://arxiv.org/abs/2401.15247}{{\ttfamily 2401.15247}}.

\bibitem{Shalgar:2022rjj}
S.~Shalgar and I.~Tamborra, \emph{{Neutrino decoupling is altered by flavor
  conversion}}, \href{https://doi.org/10.1103/PhysRevD.108.043006}{\emph{Phys.
  Rev. D} {\bfseries 108} (2023) 043006},
  [\href{https://arxiv.org/abs/2206.00676}{{\ttfamily 2206.00676}}].

\bibitem{Shalgar:2022lvv}
S.~Shalgar and I.~Tamborra, \emph{{Neutrino flavor conversion, advection, and
  collisions: Toward the full solution}},
  \href{https://doi.org/10.1103/PhysRevD.107.063025}{\emph{Phys. Rev. D}
  {\bfseries 107} (2023) 063025},
  [\href{https://arxiv.org/abs/2207.04058}{{\ttfamily 2207.04058}}].

\bibitem{Nagakura:2022xwe}
H.~Nagakura and M.~Zaizen, \emph{{Connecting small-scale to large-scale
  structures of fast neutrino-flavor conversion}},
  \href{https://doi.org/10.1103/PhysRevD.107.063033}{\emph{Phys. Rev. D}
  {\bfseries 107} (2023) 063033},
  [\href{https://arxiv.org/abs/2211.01398}{{\ttfamily 2211.01398}}].

\bibitem{Xiong:2024pue}
Z.~Xiong, M.-R. Wu, M.~George and C.-Y. Lin, \emph{{Robust integration of fast
  flavor conversions in classical neutrino transport}},
  \href{https://arxiv.org/abs/2403.17269}{{\ttfamily 2403.17269}}.

\bibitem{Cornelius:2023eop}
M.~Cornelius, S.~Shalgar and I.~Tamborra, \emph{{Perturbing fast neutrino
  flavor conversion}},
  \href{https://doi.org/10.1088/1475-7516/2024/02/038}{\emph{JCAP} {\bfseries
  02} (2024) 038}, [\href{https://arxiv.org/abs/2312.03839}{{\ttfamily
  2312.03839}}].

\bibitem{Shalgar:2019qwg}
S.~Shalgar, I.~Padilla-Gay and I.~Tamborra, \emph{{Neutrino propagation hinders
  fast pairwise flavor conversions}},
  \href{https://doi.org/10.1088/1475-7516/2020/06/048}{\emph{JCAP} {\bfseries
  06} (2020) 048}, [\href{https://arxiv.org/abs/1911.09110}{{\ttfamily
  1911.09110}}].

\bibitem{Xiong:2024tac}
Z.~Xiong, M.-R. Wu, M.~George, C.-Y. Lin, N.~K. Largani, T.~Fischer et~al.,
  \emph{{Fast neutrino flavor conversions in a supernova: emergence, evolution,
  and effects}},  \href{https://arxiv.org/abs/2402.19252}{{\ttfamily
  2402.19252}}.

\bibitem{Martin:2021xyl}
J.~D. Martin, J.~Carlson, V.~Cirigliano and H.~Duan, \emph{{Fast flavor
  oscillations in dense neutrino media with collisions}},
  \href{https://doi.org/10.1103/PhysRevD.103.063001}{\emph{Phys. Rev. D}
  {\bfseries 103} (2021) 063001},
  [\href{https://arxiv.org/abs/2101.01278}{{\ttfamily 2101.01278}}].

\bibitem{Zaizen:2023ihz}
M.~Zaizen and H.~Nagakura, \emph{{Characterizing quasisteady states of fast
  neutrino-flavor conversion by stability and conservation laws}},
  \href{https://doi.org/10.1103/PhysRevD.107.123021}{\emph{Phys. Rev. D}
  {\bfseries 107} (2023) 123021},
  [\href{https://arxiv.org/abs/2304.05044}{{\ttfamily 2304.05044}}].

\bibitem{Zaizen:2022cik}
M.~Zaizen and H.~Nagakura, \emph{{Simple method for determining asymptotic
  states of fast neutrino-flavor conversion}},
  \href{https://doi.org/10.1103/PhysRevD.107.103022}{\emph{Phys. Rev. D}
  {\bfseries 107} (2023) 103022},
  [\href{https://arxiv.org/abs/2211.09343}{{\ttfamily 2211.09343}}].

\bibitem{Xiong:2023vcm}
Z.~Xiong, M.-R. Wu, S.~Abbar, S.~Bhattacharyya, M.~George and C.-Y. Lin,
  \emph{{Evaluating approximate asymptotic distributions for fast neutrino
  flavor conversions in a periodic 1D box}},
  \href{https://doi.org/10.1103/PhysRevD.108.063003}{\emph{Phys. Rev. D}
  {\bfseries 108} (2023) 063003},
  [\href{https://arxiv.org/abs/2307.11129}{{\ttfamily 2307.11129}}].

\bibitem{Grohs:2022fyq}
E.~Grohs, S.~Richers, S.~M. Couch, F.~Foucart, J.~P. Kneller and G.~C.
  McLaughlin, \emph{{Neutrino fast flavor instability in three dimensions for a
  neutron star merger}},
  \href{https://doi.org/10.1016/j.physletb.2023.138210}{\emph{Phys. Lett. B}
  {\bfseries 846} (2023) 138210},
  [\href{https://arxiv.org/abs/2207.02214}{{\ttfamily 2207.02214}}].

\bibitem{Richers:2021xtf}
S.~Richers, D.~Willcox and N.~Ford, \emph{{Neutrino fast flavor instability in
  three dimensions}},
  \href{https://doi.org/10.1103/PhysRevD.104.103023}{\emph{Phys. Rev. D}
  {\bfseries 104} (2021) 103023},
  [\href{https://arxiv.org/abs/2109.08631}{{\ttfamily 2109.08631}}].

\bibitem{Richers:2022bkd}
S.~Richers, H.~Duan, M.-R. Wu, S.~Bhattacharyya, M.~Zaizen, M.~George et~al.,
  \emph{{Code comparison for fast flavor instability simulations}},
  \href{https://doi.org/10.1103/PhysRevD.106.043011}{\emph{Phys. Rev. D}
  {\bfseries 106} (2022) 043011},
  [\href{https://arxiv.org/abs/2205.06282}{{\ttfamily 2205.06282}}].

\bibitem{Bhattacharyya:2020jpj}
S.~Bhattacharyya and B.~Dasgupta, \emph{{Fast Flavor Depolarization of
  Supernova Neutrinos}},
  \href{https://doi.org/10.1103/PhysRevLett.126.061302}{\emph{Phys. Rev. Lett.}
  {\bfseries 126} (2021) 061302},
  [\href{https://arxiv.org/abs/2009.03337}{{\ttfamily 2009.03337}}].

\bibitem{Bhattacharyya:2022eed}
S.~Bhattacharyya and B.~Dasgupta, \emph{{Elaborating the ultimate fate of fast
  collective neutrino flavor oscillations}},
  \href{https://doi.org/10.1103/PhysRevD.106.103039}{\emph{Phys. Rev. D}
  {\bfseries 106} (2022) 103039},
  [\href{https://arxiv.org/abs/2205.05129}{{\ttfamily 2205.05129}}].

\bibitem{Wu:2021uvt}
M.-R. Wu, M.~George, C.-Y. Lin and Z.~Xiong, \emph{{Collective fast neutrino
  flavor conversions in a 1D box: Initial conditions and long-term evolution}},
  \href{https://doi.org/10.1103/PhysRevD.104.103003}{\emph{Phys. Rev. D}
  {\bfseries 104} (2021) 103003},
  [\href{https://arxiv.org/abs/2108.09886}{{\ttfamily 2108.09886}}].

\bibitem{SNarchive}
``Garching core-collapse supernova data archive.''
  \url{https://wwwmpa.mpa-garching.mpg.de/ccsnarchive/}
  \url{https://wwwmpa.mpa-garching.mpg.de/ccsnarchive/data/Bollig2016/}.

\bibitem{Sigl:1992fn}
G.~Sigl and G.~G. Raffelt, \emph{{General kinetic description of relativistic
  mixed neutrinos}},
  \href{https://doi.org/10.1016/0550-3213(93)90175-O}{\emph{Nucl. Phys. B}
  {\bfseries 406} (1993) 423--451}.

\bibitem{Esteban-Pretel:2008ovd}
A.~Esteban-Pretel, A.~Mirizzi, S.~Pastor, R.~Tom{\`a}s, G.~G. Raffelt, P.~D.
  Serpico et~al., \emph{{Role of dense matter in collective supernova neutrino
  transformations}},
  \href{https://doi.org/10.1103/PhysRevD.78.085012}{\emph{Phys. Rev. D}
  {\bfseries 78} (2008) 085012},
  [\href{https://arxiv.org/abs/0807.0659}{{\ttfamily 0807.0659}}].

\bibitem{Dasgupta:2015iia}
B.~Dasgupta and A.~Mirizzi, \emph{{Temporal Instability Enables Neutrino Flavor
  Conversions Deep Inside Supernovae}},
  \href{https://doi.org/10.1103/PhysRevD.92.125030}{\emph{Phys. Rev. D}
  {\bfseries 92} (2015) 125030},
  [\href{https://arxiv.org/abs/1509.03171}{{\ttfamily 1509.03171}}].

\bibitem{Abbar:2015fwa}
S.~Abbar and H.~Duan, \emph{{Neutrino flavor instabilities in a time-dependent
  supernova model}},
  \href{https://doi.org/10.1016/j.physletb.2015.10.019}{\emph{Phys. Lett. B}
  {\bfseries 751} (2015) 43--47},
  [\href{https://arxiv.org/abs/1509.01538}{{\ttfamily 1509.01538}}].

\bibitem{Sigl:2021tmj}
G.~Sigl, \emph{{Simulations of fast neutrino flavor conversions with
  interactions in inhomogeneous media}},
  \href{https://doi.org/10.1103/PhysRevD.105.043005}{\emph{Phys. Rev. D}
  {\bfseries 105} (2022) 043005},
  [\href{https://arxiv.org/abs/2109.00091}{{\ttfamily 2109.00091}}].

\bibitem{Nagakura:2022kic}
H.~Nagakura and M.~Zaizen, \emph{{Time-Dependent and Quasisteady Features of
  Fast Neutrino-Flavor Conversion}},
  \href{https://doi.org/10.1103/PhysRevLett.129.261101}{\emph{Phys. Rev. Lett.}
  {\bfseries 129} (2022) 261101},
  [\href{https://arxiv.org/abs/2206.04097}{{\ttfamily 2206.04097}}].

\bibitem{Padilla-Gay:2020uxa}
I.~Padilla-Gay, S.~Shalgar and I.~Tamborra, \emph{{Multi-Dimensional Solution
  of Fast Neutrino Conversions in Binary Neutron Star Merger Remnants}},
  \href{https://doi.org/10.1088/1475-7516/2021/01/017}{\emph{JCAP} {\bfseries
  01} (2021) 017}, [\href{https://arxiv.org/abs/2009.01843}{{\ttfamily
  2009.01843}}].

\bibitem{1990Ap&SS.165...65R}
M.~A. {Rudzskii}, \emph{{Kinetic equations for neutrino spin- and
  type-oscillations in a medium}},
  \href{https://doi.org/10.1007/BF00653658}{\emph{Astrophys.~and Space Science}
  {\bfseries 165} (Mar., 1990) 65--81}.

\bibitem{Janka:2012wk}
H.-T. Janka, \emph{{Explosion Mechanisms of Core-Collapse Supernovae}},
  \href{https://doi.org/10.1146/annurev-nucl-102711-094901}{\emph{Ann. Rev.
  Nucl. Part. Sci.} {\bfseries 62} (2012) 407--451},
  [\href{https://arxiv.org/abs/1206.2503}{{\ttfamily 1206.2503}}].

\bibitem{thompsonthesis}
T.~A. Thompson, \emph{Topics in the theory of core-collapse supernovae}, Ph.D.
  thesis, The University of Arizona., 2002.

\bibitem{OConnor:2014sgn}
E.~O'Connor, \emph{{An Open-Source Neutrino Radiation Hydrodynamics Code for
  Core-Collapse Supernovae}},
  \href{https://doi.org/10.1088/0067-0049/219/2/24}{\emph{Astrophys. J. Suppl.}
  {\bfseries 219} (2015) 24},
  [\href{https://arxiv.org/abs/1411.7058}{{\ttfamily 1411.7058}}].

\bibitem{Fiorillo:2024qbl}
D.~F.~G. Fiorillo and G.~Raffelt, \emph{{Fast flavor conversions at the edge of
  instability}},  \href{https://arxiv.org/abs/2403.12189}{{\ttfamily
  2403.12189}}.

\bibitem{2004cgps.book.....W}
A.~{Weiss}, W.~{Hillebrandt}, H.~C. {Thomas} and H.~{Ritter}, \emph{{Cox and
  Giuli's Principles of Stellar Structure}}.
\newblock Cambridge Scientific Publishers LTD, Cambridge, UK, 2006.

\bibitem{Morinaga:2018aug}
T.~Morinaga and S.~Yamada, \emph{{Linear stability analysis of collective
  neutrino oscillations without spurious modes}},
  \href{https://doi.org/10.1103/PhysRevD.97.023024}{\emph{Phys. Rev. D}
  {\bfseries 97} (2018) 023024},
  [\href{https://arxiv.org/abs/1803.05913}{{\ttfamily 1803.05913}}].

\bibitem{Sarikas:2012vb}
S.~Sarikas, I.~Tamborra, G.~G. Raffelt, L.~Huedepohl and H.-T. Janka,
  \emph{{Supernova neutrino halo and the suppression of self-induced flavor
  conversion}}, \href{https://doi.org/10.1103/PhysRevD.85.113007}{\emph{Phys.
  Rev. D} {\bfseries 85} (2012) 113007},
  [\href{https://arxiv.org/abs/1204.0971}{{\ttfamily 1204.0971}}].

\bibitem{Saviano:2012yh}
N.~Saviano, S.~Chakraborty, T.~Fischer and A.~Mirizzi, \emph{{Stability
  analysis of collective neutrino oscillations in the supernova accretion phase
  with realistic energy and angle distributions}},
  \href{https://doi.org/10.1103/PhysRevD.85.113002}{\emph{Phys. Rev. D}
  {\bfseries 85} (2012) 113002},
  [\href{https://arxiv.org/abs/1203.1484}{{\ttfamily 1203.1484}}].

\bibitem{Chakraborty:2011nf}
S.~Chakraborty, T.~Fischer, A.~Mirizzi, N.~Saviano and R.~Tom{\`a}s, \emph{{No
  collective neutrino flavor conversions during the supernova accretion
  phase}}, \href{https://doi.org/10.1103/PhysRevLett.107.151101}{\emph{Phys.
  Rev. Lett.} {\bfseries 107} (2011) 151101},
  [\href{https://arxiv.org/abs/1104.4031}{{\ttfamily 1104.4031}}].

\bibitem{Chakraborty:2011gd}
S.~Chakraborty, T.~Fischer, A.~Mirizzi, N.~Saviano and R.~Tom{\`a}s,
  \emph{{Analysis of matter suppression in collective neutrino oscillations
  during the supernova accretion phase}},
  \href{https://doi.org/10.1103/PhysRevD.84.025002}{\emph{Phys. Rev. D}
  {\bfseries 84} (2011) 025002},
  [\href{https://arxiv.org/abs/1105.1130}{{\ttfamily 1105.1130}}].

\bibitem{Chakraborty:2014lsa}
S.~Chakraborty, G.~Raffelt, H.-T. Janka and B.~M\"uller, \emph{{Supernova
  deleptonization asymmetry: Impact on self-induced flavor conversion}},
  \href{https://doi.org/10.1103/PhysRevD.92.105002}{\emph{Phys. Rev. D}
  {\bfseries 92} (2015) 105002},
  [\href{https://arxiv.org/abs/1412.0670}{{\ttfamily 1412.0670}}].

\bibitem{DedinNeto:2023ykt}
P.~Dedin~Neto, I.~Tamborra and S.~Shalgar, \emph{{Fast Conversion of Neutrinos:
  Energy Dependence of Flavor Instabilities}},
  \href{https://arxiv.org/abs/2312.06556}{{\ttfamily 2312.06556}}.

\bibitem{Tamborra:2017ubu}
I.~Tamborra, L.~Huedepohl, G.~Raffelt and H.-T. Janka, \emph{{Flavor-dependent
  neutrino angular distribution in core-collapse supernovae}},
  \href{https://doi.org/10.3847/1538-4357/aa6a18}{\emph{Astrophys. J.}
  {\bfseries 839} (2017) 132},
  [\href{https://arxiv.org/abs/1702.00060}{{\ttfamily 1702.00060}}].

\bibitem{Wu:2017drk}
M.-R. Wu, I.~Tamborra, O.~Just and H.-T. Janka, \emph{{Imprints of
  neutrino-pair flavor conversions on nucleosynthesis in ejecta from
  neutron-star merger remnants}},
  \href{https://doi.org/10.1103/PhysRevD.96.123015}{\emph{Phys. Rev. D}
  {\bfseries 96} (2017) 123015},
  [\href{https://arxiv.org/abs/1711.00477}{{\ttfamily 1711.00477}}].

\bibitem{Tamborra:2014aua}
I.~Tamborra, F.~Hanke, H.-T. Janka, B.~M{\"u}ller, G.~G. Raffelt and A.~Marek,
  \emph{{Self-sustained asymmetry of lepton-number emission: A new phenomenon
  during the supernova shock-accretion phase in three dimensions}},
  \href{https://doi.org/10.1088/0004-637X/792/2/96}{\emph{Astrophys. J.}
  {\bfseries 792} (2014) 96},
  [\href{https://arxiv.org/abs/1402.5418}{{\ttfamily 1402.5418}}].

\bibitem{Akaho:2023brj}
R.~Akaho, J.~Liu, H.~Nagakura, M.~Zaizen and S.~Yamada, \emph{{Collisional and
  fast neutrino flavor instabilities in two-dimensional core-collapse supernova
  simulation with Boltzmann neutrino transport}},
  \href{https://doi.org/10.1103/PhysRevD.109.023012}{\emph{Phys. Rev. D}
  {\bfseries 109} (2024) 023012},
  [\href{https://arxiv.org/abs/2311.11272}{{\ttfamily 2311.11272}}].

\bibitem{Kato:2023dcw}
C.~Kato, H.~Nagakura and M.~Zaizen, \emph{{Flavor conversions with
  energy-dependent neutrino emission and absorption}},
  \href{https://doi.org/10.1103/PhysRevD.108.023006}{\emph{Phys. Rev. D}
  {\bfseries 108} (2023) 023006},
  [\href{https://arxiv.org/abs/2303.16453}{{\ttfamily 2303.16453}}].

\bibitem{Abbar:2024ynh}
S.~Abbar, M.-R. Wu and Z.~Xiong, \emph{{Application of neural networks for the
  reconstruction of supernova neutrino energy spectra following fast neutrino
  flavor conversions}},
  \href{https://doi.org/10.1103/PhysRevD.109.083019}{\emph{Phys. Rev. D}
  {\bfseries 109} (2024) 083019},
  [\href{https://arxiv.org/abs/2401.17424}{{\ttfamily 2401.17424}}].

\bibitem{Nagakura:2023jfi}
H.~Nagakura, L.~Johns and M.~Zaizen, \emph{{Bhatnagar-Gross-Krook subgrid model
  for neutrino quantum kinetics}},
  \href{https://doi.org/10.1103/PhysRevD.109.083013}{\emph{Phys. Rev. D}
  {\bfseries 109} (2024) 083013},
  [\href{https://arxiv.org/abs/2312.16285}{{\ttfamily 2312.16285}}].

\bibitem{Abbar:2023ltx}
S.~Abbar, M.-R. Wu and Z.~Xiong, \emph{{Physics-informed neural networks for
  predicting the asymptotic outcome of fast neutrino flavor conversions}},
  \href{https://doi.org/10.1103/PhysRevD.109.043024}{\emph{Phys. Rev. D}
  {\bfseries 109} (2024) 043024},
  [\href{https://arxiv.org/abs/2311.15656}{{\ttfamily 2311.15656}}].

\end{thebibliography}\endgroup
\end{document}